%\input jytex.tex   % available from hep-th
% File jytex.tex, for jyTeX version 2.6M (June 1992)
% Copyright (c) 1991, 1992 by Jonathan P. Yamron
% For full documentation, "get jydoc" from hep-ph@xxx.lanl.gov
%   Problems?  Contact brahm@theory3.caltech.edu.

\catcode`\@=11

%*****************************************************************************

\message{Loading jyTeX fonts...}

%************************************************************
%*
%*             Available fonts
%*
%************************************************************

%************** 5-point fonts *******************************

\font\vptrm=cmr5
\font\vptmit=cmmi5
\font\vptsy=cmsy5
\font\vptbf=cmbx5

\skewchar\vptmit='177 \skewchar\vptsy='60
\fontdimen16 \vptsy=\the\fontdimen17 \vptsy

\def\vpt{\ifmmode\err@badsizechange\else
     \@mathfontinit
     \textfont0=\vptrm  \scriptfont0=\vptrm  \scriptscriptfont0=\vptrm
     \textfont1=\vptmit \scriptfont1=\vptmit \scriptscriptfont1=\vptmit
     \textfont2=\vptsy  \scriptfont2=\vptsy  \scriptscriptfont2=\vptsy
     \textfont3=\xptex  \scriptfont3=\xptex  \scriptscriptfont3=\xptex
     \textfont\bffam=\vptbf
     \scriptfont\bffam=\vptbf
     \scriptscriptfont\bffam=\vptbf
     \@fontstyleinit
     \def\rm{\vptrm\fam=\z@}%
     \def\bf{\vptbf\fam=\bffam}%
     \def\oldstyle{\vptmit\fam=\@ne}%
     \rm\fi}

%************** 6-point fonts *******************************

\font\viptrm=cmr6
\font\viptmit=cmmi6
\font\viptsy=cmsy6
\font\viptbf=cmbx6

\skewchar\viptmit='177 \skewchar\viptsy='60
\fontdimen16 \viptsy=\the\fontdimen17 \viptsy

\def\vipt{\ifmmode\err@badsizechange\else
     \@mathfontinit
     \textfont0=\viptrm  \scriptfont0=\vptrm  \scriptscriptfont0=\vptrm
     \textfont1=\viptmit \scriptfont1=\vptmit \scriptscriptfont1=\vptmit
     \textfont2=\viptsy  \scriptfont2=\vptsy  \scriptscriptfont2=\vptsy
     \textfont3=\xptex   \scriptfont3=\xptex  \scriptscriptfont3=\xptex
     \textfont\bffam=\viptbf
     \scriptfont\bffam=\vptbf
     \scriptscriptfont\bffam=\vptbf
     \@fontstyleinit
     \def\rm{\viptrm\fam=\z@}%
     \def\bf{\viptbf\fam=\bffam}%
     \def\oldstyle{\viptmit\fam=\@ne}%
     \rm\fi}

%************** 7-point fonts *******************************

\font\viiptrm=cmr7
\font\viiptmit=cmmi7
\font\viiptsy=cmsy7
\font\viiptit=cmti7
\font\viiptbf=cmbx7

\skewchar\viiptmit='177 \skewchar\viiptsy='60
\fontdimen16 \viiptsy=\the\fontdimen17 \viiptsy

\def\viipt{\ifmmode\err@badsizechange\else
     \@mathfontinit
     \textfont0=\viiptrm  \scriptfont0=\vptrm  \scriptscriptfont0=\vptrm
     \textfont1=\viiptmit \scriptfont1=\vptmit \scriptscriptfont1=\vptmit
     \textfont2=\viiptsy  \scriptfont2=\vptsy  \scriptscriptfont2=\vptsy
     \textfont3=\xptex    \scriptfont3=\xptex  \scriptscriptfont3=\xptex
     \textfont\itfam=\viiptit
     \scriptfont\itfam=\viiptit
     \scriptscriptfont\itfam=\viiptit
     \textfont\bffam=\viiptbf
     \scriptfont\bffam=\vptbf
     \scriptscriptfont\bffam=\vptbf
     \@fontstyleinit
     \def\rm{\viiptrm\fam=\z@}%
     \def\it{\viiptit\fam=\itfam}%
     \def\bf{\viiptbf\fam=\bffam}%
     \def\oldstyle{\viiptmit\fam=\@ne}%
     \rm\fi}

%************** 8-point fonts *******************************

\font\viiiptrm=cmr8
\font\viiiptmit=cmmi8
\font\viiiptsy=cmsy8
\font\viiiptit=cmti8
%\font\viiiptsl=cmsl8
\font\viiiptbf=cmbx8
%\font\viiipttt=cmtt8
%\font\viiiptss=cmss8

\skewchar\viiiptmit='177 \skewchar\viiiptsy='60
\fontdimen16 \viiiptsy=\the\fontdimen17 \viiiptsy

\def\viiipt{\ifmmode\err@badsizechange\else
     \@mathfontinit
     \textfont0=\viiiptrm  \scriptfont0=\viptrm  \scriptscriptfont0=\vptrm
     \textfont1=\viiiptmit \scriptfont1=\viptmit \scriptscriptfont1=\vptmit
     \textfont2=\viiiptsy  \scriptfont2=\viptsy  \scriptscriptfont2=\vptsy
     \textfont3=\xptex     \scriptfont3=\xptex   \scriptscriptfont3=\xptex
     \textfont\itfam=\viiiptit
     \scriptfont\itfam=\viiptit
     \scriptscriptfont\itfam=\viiptit
     \textfont\bffam=\viiiptbf
     \scriptfont\bffam=\viptbf
     \scriptscriptfont\bffam=\vptbf
     \@fontstyleinit
     \def\rm{\viiiptrm\fam=\z@}%
     \def\it{\viiiptit\fam=\itfam}%
     \def\bf{\viiiptbf\fam=\bffam}%
     \def\oldstyle{\viiiptmit\fam=\@ne}%
     \rm\fi}

%************** Optional 9-point fonts **********************

\def\getixpt{%
     \font\ixptrm=cmr9
     \font\ixptmit=cmmi9
     \font\ixptsy=cmsy9
     \font\ixptit=cmti9
%     \font\ixptsl=cmsl9
     \font\ixptbf=cmbx9
%     \font\ixpttt=cmtt9
%     \font\ixptss=cmss9
     \skewchar\ixptmit='177 \skewchar\ixptsy='60
     \fontdimen16 \ixptsy=\the\fontdimen17 \ixptsy}

\def\ixpt{\ifmmode\err@badsizechange\else
     \@mathfontinit
     \textfont0=\ixptrm  \scriptfont0=\viiptrm  \scriptscriptfont0=\vptrm
     \textfont1=\ixptmit \scriptfont1=\viiptmit \scriptscriptfont1=\vptmit
     \textfont2=\ixptsy  \scriptfont2=\viiptsy  \scriptscriptfont2=\vptsy
     \textfont3=\xptex   \scriptfont3=\xptex    \scriptscriptfont3=\xptex
     \textfont\itfam=\ixptit
     \scriptfont\itfam=\viiptit
     \scriptscriptfont\itfam=\viiptit
     \textfont\bffam=\ixptbf
     \scriptfont\bffam=\viiptbf
     \scriptscriptfont\bffam=\vptbf
     \@fontstyleinit
     \def\rm{\ixptrm\fam=\z@}%
     \def\it{\ixptit\fam=\itfam}%
     \def\bf{\ixptbf\fam=\bffam}%
     \def\oldstyle{\ixptmit\fam=\@ne}%
     \rm\fi}

%************** 10-point fonts ******************************

\font\xptrm=cmr10
\font\xptmit=cmmi10
\font\xptsy=cmsy10
\font\xptex=cmex10
\font\xptit=cmti10
\font\xptsl=cmsl10
\font\xptbf=cmbx10
\font\xpttt=cmtt10
\font\xptss=cmss10
\font\xptsc=cmcsc10
\font\xptbfs=cmb10
\font\xptbmit=cmmib10

\skewchar\xptmit='177 \skewchar\xptbmit='177 \skewchar\xptsy='60
\fontdimen16 \xptsy=\the\fontdimen17 \xptsy

\def\xpt{\ifmmode\err@badsizechange\else
     \@mathfontinit
     \textfont0=\xptrm  \scriptfont0=\viiptrm  \scriptscriptfont0=\vptrm
     \textfont1=\xptmit \scriptfont1=\viiptmit \scriptscriptfont1=\vptmit
     \textfont2=\xptsy  \scriptfont2=\viiptsy  \scriptscriptfont2=\vptsy
     \textfont3=\xptex  \scriptfont3=\xptex    \scriptscriptfont3=\xptex
     \textfont\itfam=\xptit
     \scriptfont\itfam=\viiptit
     \scriptscriptfont\itfam=\viiptit
     \textfont\bffam=\xptbf
     \scriptfont\bffam=\viiptbf
     \scriptscriptfont\bffam=\vptbf
     \textfont\bfsfam=\xptbfs
     \scriptfont\bfsfam=\viiptbf
     \scriptscriptfont\bfsfam=\vptbf
     \textfont\bmitfam=\xptbmit
     \scriptfont\bmitfam=\viiptmit
     \scriptscriptfont\bmitfam=\vptmit
     \@fontstyleinit
     \def\rm{\xptrm\fam=\z@}%
     \def\it{\xptit\fam=\itfam}%
     \def\sl{\xptsl}%
     \def\bf{\xptbf\fam=\bffam}%
     \def\tt{\xpttt}%
     \def\ss{\xptss}%
     \def\sc{\xptsc}%
     \def\bfs{\xptbfs\fam=\bfsfam}%
     \def\bmit{\fam=\bmitfam}%
     \def\oldstyle{\xptmit\fam=\@ne}%
     \rm\fi}

%************** Optional 11-point fonts *********************

\def\getxipt{%
     \font\xiptrm=cmr10  scaled\magstephalf
     \font\xiptmit=cmmi10 scaled\magstephalf
     \font\xiptsy=cmsy10 scaled\magstephalf
     \font\xiptex=cmex10 scaled\magstephalf
     \font\xiptit=cmti10 scaled\magstephalf
     \font\xiptsl=cmsl10 scaled\magstephalf
     \font\xiptbf=cmbx10 scaled\magstephalf
     \font\xipttt=cmtt10 scaled\magstephalf
     \font\xiptss=cmss10 scaled\magstephalf
     \skewchar\xiptmit='177 \skewchar\xiptsy='60
     \fontdimen16 \xiptsy=\the\fontdimen17 \xiptsy}

\def\xipt{\ifmmode\err@badsizechange\else
     \@mathfontinit
     \textfont0=\xiptrm  \scriptfont0=\viiiptrm  \scriptscriptfont0=\viptrm
     \textfont1=\xiptmit \scriptfont1=\viiiptmit \scriptscriptfont1=\viptmit
     \textfont2=\xiptsy  \scriptfont2=\viiiptsy  \scriptscriptfont2=\viptsy
     \textfont3=\xiptex  \scriptfont3=\xptex     \scriptscriptfont3=\xptex
     \textfont\itfam=\xiptit
     \scriptfont\itfam=\viiiptit
     \scriptscriptfont\itfam=\viiptit
     \textfont\bffam=\xiptbf
     \scriptfont\bffam=\viiiptbf
     \scriptscriptfont\bffam=\viptbf
     \@fontstyleinit
     \def\rm{\xiptrm\fam=\z@}%
     \def\it{\xiptit\fam=\itfam}%
     \def\sl{\xiptsl}%
     \def\bf{\xiptbf\fam=\bffam}%
     \def\tt{\xipttt}%
     \def\ss{\xiptss}%
     \def\oldstyle{\xiptmit\fam=\@ne}%
     \rm\fi}

%************** 12-point fonts ******************************

\font\xiiptrm=cmr12
\font\xiiptmit=cmmi12
\font\xiiptsy=cmsy10  scaled\magstep1
\font\xiiptex=cmex10  scaled\magstep1
\font\xiiptit=cmti12
\font\xiiptsl=cmsl12
\font\xiiptbf=cmbx12
\font\xiipttt=cmtt12
\font\xiiptss=cmss12
\font\xiiptsc=cmcsc10 scaled\magstep1
\font\xiiptbfs=cmb10  scaled\magstep1
\font\xiiptbmit=cmmib10 scaled\magstep1

\skewchar\xiiptmit='177 \skewchar\xiiptbmit='177 \skewchar\xiiptsy='60
\fontdimen16 \xiiptsy=\the\fontdimen17 \xiiptsy

\def\xiipt{\ifmmode\err@badsizechange\else
     \@mathfontinit
     \textfont0=\xiiptrm  \scriptfont0=\viiiptrm  \scriptscriptfont0=\viptrm
     \textfont1=\xiiptmit \scriptfont1=\viiiptmit \scriptscriptfont1=\viptmit
     \textfont2=\xiiptsy  \scriptfont2=\viiiptsy  \scriptscriptfont2=\viptsy
     \textfont3=\xiiptex  \scriptfont3=\xptex     \scriptscriptfont3=\xptex
     \textfont\itfam=\xiiptit
     \scriptfont\itfam=\viiiptit
     \scriptscriptfont\itfam=\viiptit
     \textfont\bffam=\xiiptbf
     \scriptfont\bffam=\viiiptbf
     \scriptscriptfont\bffam=\viptbf
     \textfont\bfsfam=\xiiptbfs
     \scriptfont\bfsfam=\viiiptbf
     \scriptscriptfont\bfsfam=\viptbf
     \textfont\bmitfam=\xiiptbmit
     \scriptfont\bmitfam=\viiiptmit
     \scriptscriptfont\bmitfam=\viptmit
     \@fontstyleinit
     \def\rm{\xiiptrm\fam=\z@}%
     \def\it{\xiiptit\fam=\itfam}%
     \def\sl{\xiiptsl}%
     \def\bf{\xiiptbf\fam=\bffam}%
     \def\tt{\xiipttt}%
     \def\ss{\xiiptss}%
     \def\sc{\xiiptsc}%
     \def\bfs{\xiiptbfs\fam=\bfsfam}%
     \def\bmit{\fam=\bmitfam}%
     \def\oldstyle{\xiiptmit\fam=\@ne}%
     \rm\fi}

%************** Optional 13-point fonts *********************

\def\getxiiipt{%
     \font\xiiiptrm=cmr12  scaled\magstephalf
     \font\xiiiptmit=cmmi12 scaled\magstephalf
     \font\xiiiptsy=cmsy9  scaled\magstep2
     \font\xiiiptit=cmti12 scaled\magstephalf
     \font\xiiiptsl=cmsl12 scaled\magstephalf
     \font\xiiiptbf=cmbx12 scaled\magstephalf
     \font\xiiipttt=cmtt12 scaled\magstephalf
     \font\xiiiptss=cmss12 scaled\magstephalf
     \skewchar\xiiiptmit='177 \skewchar\xiiiptsy='60
     \fontdimen16 \xiiiptsy=\the\fontdimen17 \xiiiptsy}

\def\xiiipt{\ifmmode\err@badsizechange\else
     \@mathfontinit
     \textfont0=\xiiiptrm  \scriptfont0=\xptrm  \scriptscriptfont0=\viiptrm
     \textfont1=\xiiiptmit \scriptfont1=\xptmit \scriptscriptfont1=\viiptmit
     \textfont2=\xiiiptsy  \scriptfont2=\xptsy  \scriptscriptfont2=\viiptsy
     \textfont3=\xivptex   \scriptfont3=\xptex  \scriptscriptfont3=\xptex
     \textfont\itfam=\xiiiptit
     \scriptfont\itfam=\xptit
     \scriptscriptfont\itfam=\viiptit
     \textfont\bffam=\xiiiptbf
     \scriptfont\bffam=\xptbf
     \scriptscriptfont\bffam=\viiptbf
     \@fontstyleinit
     \def\rm{\xiiiptrm\fam=\z@}%
     \def\it{\xiiiptit\fam=\itfam}%
     \def\sl{\xiiiptsl}%
     \def\bf{\xiiiptbf\fam=\bffam}%
     \def\tt{\xiiipttt}%
     \def\ss{\xiiiptss}%
     \def\oldstyle{\xiiiptmit\fam=\@ne}%
     \rm\fi}

%************** 14-point fonts ******************************

\font\xivptrm=cmr12   scaled\magstep1
\font\xivptmit=cmmi12  scaled\magstep1
\font\xivptsy=cmsy10  scaled\magstep2
\font\xivptex=cmex10  scaled\magstep2
\font\xivptit=cmti12  scaled\magstep1
\font\xivptsl=cmsl12  scaled\magstep1
\font\xivptbf=cmbx12  scaled\magstep1
\font\xivpttt=cmtt12  scaled\magstep1
\font\xivptss=cmss12  scaled\magstep1
\font\xivptsc=cmcsc10 scaled\magstep2
\font\xivptbfs=cmb10  scaled\magstep2
\font\xivptbmit=cmmib10 scaled\magstep2

\skewchar\xivptmit='177 \skewchar\xivptbmit='177 \skewchar\xivptsy='60
\fontdimen16 \xivptsy=\the\fontdimen17 \xivptsy

\def\xivpt{\ifmmode\err@badsizechange\else
     \@mathfontinit
     \textfont0=\xivptrm  \scriptfont0=\xptrm  \scriptscriptfont0=\viiptrm
     \textfont1=\xivptmit \scriptfont1=\xptmit \scriptscriptfont1=\viiptmit
     \textfont2=\xivptsy  \scriptfont2=\xptsy  \scriptscriptfont2=\viiptsy
     \textfont3=\xivptex  \scriptfont3=\xptex  \scriptscriptfont3=\xptex
     \textfont\itfam=\xivptit
     \scriptfont\itfam=\xptit
     \scriptscriptfont\itfam=\viiptit
     \textfont\bffam=\xivptbf
     \scriptfont\bffam=\xptbf
     \scriptscriptfont\bffam=\viiptbf
     \textfont\bfsfam=\xivptbfs
     \scriptfont\bfsfam=\xptbfs
     \scriptscriptfont\bfsfam=\viiptbf
     \textfont\bmitfam=\xivptbmit
     \scriptfont\bmitfam=\xptbmit
     \scriptscriptfont\bmitfam=\viiptmit
     \@fontstyleinit
     \def\rm{\xivptrm\fam=\z@}%
     \def\it{\xivptit\fam=\itfam}%
     \def\sl{\xivptsl}%
     \def\bf{\xivptbf\fam=\bffam}%
     \def\tt{\xivpttt}%
     \def\ss{\xivptss}%
     \def\sc{\xivptsc}%
     \def\bfs{\xivptbfs\fam=\bfsfam}%
     \def\bmit{\fam=\bmitfam}%
     \def\oldstyle{\xivptmit\fam=\@ne}%
     \rm\fi}

%************** 17-point fonts ******************************

\font\xviiptrm=cmr17
\font\xviiptmit=cmmi12 scaled\magstep2
\font\xviiptsy=cmsy10 scaled\magstep3
\font\xviiptex=cmex10 scaled\magstep3
\font\xviiptit=cmti12 scaled\magstep2
\font\xviiptbf=cmbx12 scaled\magstep2
\font\xviiptbfs=cmb10 scaled\magstep3

\skewchar\xviiptmit='177 \skewchar\xviiptsy='60
\fontdimen16 \xviiptsy=\the\fontdimen17 \xviiptsy

\def\xviipt{\ifmmode\err@badsizechange\else
     \@mathfontinit
     \textfont0=\xviiptrm  \scriptfont0=\xiiptrm  \scriptscriptfont0=\viiiptrm
     \textfont1=\xviiptmit \scriptfont1=\xiiptmit \scriptscriptfont1=\viiiptmit
     \textfont2=\xviiptsy  \scriptfont2=\xiiptsy  \scriptscriptfont2=\viiiptsy
     \textfont3=\xviiptex  \scriptfont3=\xiiptex  \scriptscriptfont3=\xptex
     \textfont\itfam=\xviiptit
     \scriptfont\itfam=\xiiptit
     \scriptscriptfont\itfam=\viiiptit
     \textfont\bffam=\xviiptbf
     \scriptfont\bffam=\xiiptbf
     \scriptscriptfont\bffam=\viiiptbf
     \textfont\bfsfam=\xviiptbfs
     \scriptfont\bfsfam=\xiiptbfs
     \scriptscriptfont\bfsfam=\viiiptbf
     \@fontstyleinit
     \def\rm{\xviiptrm\fam=\z@}%
     \def\it{\xviiptit\fam=\itfam}%
     \def\bf{\xviiptbf\fam=\bffam}%
     \def\bfs{\xviiptbfs\fam=\bfsfam}%
     \def\oldstyle{\xviiptmit\fam=\@ne}%
     \rm\fi}

%************** 21-point fonts ******************************

\font\xxiptrm=cmr17  scaled\magstep1
%\font\xxiptmit=cmmi12 scaled\magstep3
%\font\xxiptsy=cmsy10 scaled\magstep4
%\font\xxiptex=cmex10 scaled\magstep4
%\font\xxiptbf=cmbx12 scaled\magstep3

%\skewchar\xxiptmit='177 \skewchar\xxiptsy='60
%\fontdimen16 \xxiptsy=\the\fontdimen17 \xxiptsy

\def\xxipt{\ifmmode\err@badsizechange\else
     \@mathfontinit
%     \textfont0=\xxiptrm  \scriptfont0=\xivptrm  \scriptscriptfont0=\xptrm
%     \textfont1=\xxiptmit \scriptfont1=\xivptmit \scriptscriptfont1=\xptmit
%     \textfont2=\xxiptsy  \scriptfont2=\xivptsy  \scriptscriptfont2=\xptsy
%     \textfont3=\xxiptex  \scriptfont3=\xivptex  \scriptscriptfont3=\xptex
%     \textfont\bffam=\xxiptbf
%     \scriptfont\bffam=\xivptbf
%     \scriptscriptfont\bffam=\xptbf
     \@fontstyleinit
     \def\rm{\xxiptrm\fam=\z@}%
     \rm\fi}

%************** 25-point fonts ******************************

\font\xxvptrm=cmr17  scaled\magstep2
%\font\xxvptmit=cmmi12 scaled\magstep4
%\font\xxvptsy=cmsy10 scaled\magstep5
%\font\xxvptex=cmex10 scaled\magstep5
%\font\xxvptbf=cmbx12 scaled\magstep4

%\skewchar\xxvptmit='177 \skewchar\xxvptsy='60
%\fontdimen16 \xxvptsy=\the\fontdimen17 \xxvptsy

\def\xxvpt{\ifmmode\err@badsizechange\else
     \@mathfontinit
%     \textfont0=\xxvptrm  \scriptfont0=\xviiptrm  \scriptscriptfont0=\xiiptrm
%     \textfont1=\xxvptmit \scriptfont1=\xviiptmit \scriptscriptfont1=\xiiptmit
%     \textfont2=\xxvptsy  \scriptfont2=\xviiptsy  \scriptscriptfont2=\xiiptsy
%     \textfont3=\xxvptex  \scriptfont3=\xviiptex  \scriptscriptfont3=\xiiptex
%     \textfont\bffam=\xxvptbf
%     \scriptfont\bffam=\xviiptbf
%     \scriptscriptfont\bffam=\xiiptbf
     \@fontstyleinit
     \def\rm{\xxvptrm\fam=\z@}%
     \rm\fi}

%************** Other fonts *********************************

%\font\dummy=dummy

%******************************************************************************

\message{Loading jyTeX macros...}

%************************************************************
%*
%*              Simple modifications to plain
%*
%************************************************************
\message{modifications to plain.tex,}

% The "\outer" qualifier is removed from the definitions of \newcount through
% \newif so that they may be used in definitions.  \newif is also changed to
% make \if commands globally defined.

\def\newcount{\alloc@0\count\countdef\insc@unt}
\def\newdimen{\alloc@1\dimen\dimendef\insc@unt}
\def\newskip{\alloc@2\skip\skipdef\insc@unt}
\def\newmuskip{\alloc@3\muskip\muskipdef\@cclvi}
\def\newbox{\alloc@4\box\chardef\insc@unt}
\def\newtoks{\alloc@5\toks\toksdef\@cclvi}
\def\newhelp#1#2{\newtoks#1\global#1\expandafter{\csname#2\endcsname}}
\def\newread{\alloc@6\read\chardef\sixt@@n}
\def\newwrite{\alloc@7\write\chardef\sixt@@n}
\def\newfam{\alloc@8\fam\chardef\sixt@@n}
\def\newinsert#1{\global\advance\insc@unt by\m@ne
     \ch@ck0\insc@unt\count
     \ch@ck1\insc@unt\dimen
     \ch@ck2\insc@unt\skip
     \ch@ck4\insc@unt\box
     \allocationnumber=\insc@unt
     \global\chardef#1=\allocationnumber
     \wlog{\string#1=\string\insert\the\allocationnumber}}
\def\newif#1{\count@\escapechar \escapechar\m@ne
     \expandafter\expandafter\expandafter
          \xdef\@if#1{true}{\let\noexpand#1=\noexpand\iftrue}%
     \expandafter\expandafter\expandafter
          \xdef\@if#1{false}{\let\noexpand#1=\noexpand\iffalse}%
     \global\@if#1{false}\escapechar=\count@}

%************** Some parameter changes **********************

\newlinechar=`\^^J
\overfullrule=0pt

%************** Font-related modifications ******************

% The plain fonts are mapped onto the corresponding jyTeX fonts

% Some control sequences are disabled.

\let\itfam=\undefined

\let\bffam=\undefined

\count18=3

% German sharp s is given a new name (\ss is already taken)

\chardef\sharps="19

% The mathcode assignments of characters in the math italic font are changed to
% allow for switching to boldface.

\mathchardef\alpha="710B
\mathchardef\beta="710C
\mathchardef\gamma="710D
\mathchardef\delta="710E
\mathchardef\epsilon="710F
\mathchardef\zeta="7110
\mathchardef\eta="7111
\mathchardef\theta="7112
\mathchardef\iota="7113
\mathchardef\kappa="7114
\mathchardef\lambda="7115
\mathchardef\mu="7116
\mathchardef\nu="7117
\mathchardef\xi="7118
\mathchardef\pi="7119
\mathchardef\rho="711A
\mathchardef\sigma="711B
\mathchardef\tau="711C
\mathchardef\upsilon="711D
\mathchardef\phi="711E
\mathchardef\chi="711F
\mathchardef\psi="7120
\mathchardef\omega="7121
\mathchardef\varepsilon="7122
\mathchardef\vartheta="7123
\mathchardef\varpi="7124
\mathchardef\varrho="7125
\mathchardef\varsigma="7126
\mathchardef\varphi="7127
\mathchardef\imath="717B
\mathchardef\jmath="717C
\mathchardef\ell="7160
\mathchardef\wp="717D
\mathchardef\partial="7140
\mathchardef\flat="715B
\mathchardef\natural="715C
\mathchardef\sharp="715D

%************** Miscellaneous changes ***********************

% The dimension \p@ (1pt) is replaced with \rp@ (relative pt, defined below),
% whose size is determined by the base type size of the document.

\def\angle{{\vbox{\ialign{$\m@th\scriptstyle##$\crcr
     \not\mathrel{\mkern14mu}\crcr
     \noalign{\nointerlineskip}
     \mkern2.5mu\leaders\hrule height.34\rp@\hfill\mkern2.5mu\crcr}}}}
\def\vdots{\vbox{\baselineskip4\rp@ \lineskiplimit\z@
     \kern6\rp@\hbox{.}\hbox{.}\hbox{.}}}
\def\ddots{\mathinner{\mkern1mu\raise7\rp@\vbox{\kern7\rp@\hbox{.}}\mkern2mu
     \raise4\rp@\hbox{.}\mkern2mu\raise\rp@\hbox{.}\mkern1mu}}
\def\overrightarrow#1{\vbox{\ialign{##\crcr
     \rightarrowfill\crcr
     \noalign{\kern-\rp@\nointerlineskip}
     $\hfil\displaystyle{#1}\hfil$\crcr}}}
\def\overleftarrow#1{\vbox{\ialign{##\crcr
     \leftarrowfill\crcr
     \noalign{\kern-\rp@\nointerlineskip}
     $\hfil\displaystyle{#1}\hfil$\crcr}}}
\def\overbrace#1{\mathop{\vbox{\ialign{##\crcr
     \noalign{\kern3\rp@}
     \downbracefill\crcr
     \noalign{\kern3\rp@\nointerlineskip}
     $\hfil\displaystyle{#1}\hfil$\crcr}}}\limits}
\def\underbrace#1{\mathop{\vtop{\ialign{##\crcr
     $\hfil\displaystyle{#1}\hfil$\crcr
     \noalign{\kern3\rp@\nointerlineskip}
     \upbracefill\crcr
     \noalign{\kern3\rp@}}}}\limits}
\def\big#1{{\hbox{$\left#1\vbox to8.5\rp@ {}\right.\n@space$}}}
\def\Big#1{{\hbox{$\left#1\vbox to11.5\rp@ {}\right.\n@space$}}}
\def\bigg#1{{\hbox{$\left#1\vbox to14.5\rp@ {}\right.\n@space$}}}
\def\Bigg#1{{\hbox{$\left#1\vbox to17.5\rp@ {}\right.\n@space$}}}
\def\@vereq#1#2{\lower.5\rp@\vbox{\baselineskip\z@skip\lineskip-.5\rp@
     \ialign{$\m@th#1\hfil##\hfil$\crcr#2\crcr=\crcr}}}
\def\rlh@#1{\vcenter{\hbox{\ooalign{\raise2\rp@
     \hbox{$#1\rightharpoonup$}\crcr
     $#1\leftharpoondown$}}}}
\def\bordermatrix#1{\begingroup\m@th
     \setbox\z@\vbox{%
          \def\cr{\crcr\noalign{\kern2\rp@\global\let\cr\endline}}%
          \ialign{$##$\hfil\kern2\rp@\kern\p@renwd
               &\thinspace\hfil$##$\hfil&&\quad\hfil$##$\hfil\crcr
               \omit\strut\hfil\crcr
               \noalign{\kern-\baselineskip}%
               #1\crcr\omit\strut\cr}}%
     \setbox\tw@\vbox{\unvcopy\z@\global\setbox\@ne\lastbox}%
     \setbox\tw@\hbox{\unhbox\@ne\unskip\global\setbox\@ne\lastbox}%
     \setbox\tw@\hbox{$\kern\wd\@ne\kern-\p@renwd\left(\kern-\wd\@ne
          \global\setbox\@ne\vbox{\box\@ne\kern2\rp@}%
          \vcenter{\kern-\ht\@ne\unvbox\z@\kern-\baselineskip}%
          \,\right)$}%
     \null\;\vbox{\kern\ht\@ne\box\tw@}\endgroup}
\def\endinsert{\egroup
     \if@mid\dimen@\ht\z@
          \advance\dimen@\dp\z@
          \advance\dimen@12\rp@
          \advance\dimen@\pagetotal
          \ifdim\dimen@>\pagegoal\@midfalse\p@gefalse\fi
     \fi
     \if@mid\bigskip\box\z@
          \bigbreak
     \else\insert\topins{\penalty100 \splittopskip\z@skip
               \splitmaxdepth\maxdimen\floatingpenalty\z@
               \ifp@ge\dimen@\dp\z@
                    \vbox to\vsize{\unvbox\z@\kern-\dimen@}%
               \else\box\z@\nobreak\bigskip
               \fi}%
     \fi
     \endgroup}

% \normalbaselines is removed from \cases and \matrix.

\def\cases#1{\left\{\,\vcenter{\m@th
     \ialign{$##\hfil$&\quad##\hfil\crcr#1\crcr}}\right.}
\def\matrix#1{\null\,\vcenter{\m@th
     \ialign{\hfil$##$\hfil&&\quad\hfil$##$\hfil\crcr
          \mathstrut\crcr
          \noalign{\kern-\baselineskip}
          #1\crcr
          \mathstrut\crcr
          \noalign{\kern-\baselineskip}}}\,}

% \raggedbottom modified slightly

\newif\ifraggedbottom

\def\raggedbottom{\ifraggedbottom\else
     \advance\topskip by\z@ plus60pt \raggedbottomtrue\fi}%
\def\normalbottom{\ifraggedbottom
     \advance\topskip by\z@ plus-60pt \raggedbottomfalse\fi}

%************************************************************
%*
%*              Miscellaneous definitions
%*
%************************************************************
\message{hacks,}

%************** Hack registers ******************************

\toksdef\toks@i=1
\toksdef\toks@ii=2

%************** Basic macros ********************************

\def\TeX{T\kern-.1667em \lower.5ex \hbox{E}\kern-.125em X\null}
\def\jyTeX{{\leavevmode
     \raise.587ex \hbox{\it\j}\kern-.1em \lower.048ex \hbox{\it y}\kern-.12em
     \TeX}}

\let\then=\iftrue
\def\ifnoarg#1\then{\def\hack@{#1}\ifx\hack@\empty}
\def\ifundefined#1\then{%
     \expandafter\ifx\csname\expandafter\blank\string#1\endcsname\relax}
\def\useif#1\then{\csname#1\endcsname}
\def\usename#1{\csname#1\endcsname}
\def\useafter#1#2{\expandafter#1\csname#2\endcsname}

% Modify so that I can have \loop's within \loop's?
\long\def\loop#1\repeat{\def\@iterate{#1\expandafter\@iterate\fi}\@iterate
     \let\@iterate=\relax}
%\long\def\loop#1\repeat{\def\@loopbody{#1}\@iterate}
%\def\@iterate{\@loopbody\let\next=\@iterate\else\let\next=\relax\fi\next}

\let\TeXend=\end
\def\begin#1{\begingroup\def\@@blockname{#1}\usename{begin#1}}
\def\end#1{\usename{end#1}\def\hack@{#1}%
     \ifx\@@blockname\hack@
          \endgroup
     \else\err@badgroup\hack@\@@blockname
     \fi}
\def\@@blockname{}

\def\defaultoption[#1]#2{%
     \def\hack@{\ifx\hack@ii[\toks@={#2}\else\toks@={#2[#1]}\fi\the\toks@}%
     \futurelet\hack@ii\hack@}

\def\markup#1{\let\@@marksf=\empty
     \ifhmode\edef\@@marksf{\spacefactor=\the\spacefactor\relax}\/\fi
     ${}^{\hbox{\subscriptfonts#1}}$\@@marksf}

%************** Time registers ******************************

\newtoks\shortyear
\newtoks\militaryhour
\newtoks\standardhour
\newtoks\minute
\newtoks\amorpm

\def\settime{\count@=\time\divide\count@ by60
     \militaryhour=\expandafter{\number\count@}%
     {\multiply\count@ by-60 \advance\count@ by\time
          \xdef\hack@{\ifnum\count@<10 0\fi\number\count@}}%
     \minute=\expandafter{\hack@}%
     \ifnum\count@<12
          \amorpm={am}
     \else\amorpm={pm}
          \ifnum\count@>12 \advance\count@ by-12 \fi
     \fi
     \standardhour=\expandafter{\number\count@}%
     \def\hack@19##1##2{\shortyear={##1##2}}%
          \expandafter\hack@\the\year}

\def\monthword#1{%
     \ifcase#1
          $\bullet$\err@badcountervalue{monthword}%
          \or January\or February\or March\or April\or May\or June%
          \or July\or August\or September\or October\or November\or December%
     \else$\bullet$\err@badcountervalue{monthword}%
     \fi}

\def\monthabbr#1{%
     \ifcase#1
          $\bullet$\err@badcountervalue{monthabbr}%
          \or Jan\or Feb\or Mar\or Apr\or May\or Jun%
          \or Jul\or Aug\or Sep\or Oct\or Nov\or Dec%
     \else$\bullet$\err@badcountervalue{monthabbr}%
     \fi}

\def\militarytime{\the\militaryhour:\the\minute}
\def\standardtime{\the\standardhour:\the\minute}

%************** Number styles *******************************

\def\@setnumstyle#1#2{\expandafter\global\expandafter\expandafter
     \expandafter\let\expandafter\expandafter
     \csname @\expandafter\blank\string#1style\endcsname
     \csname#2\endcsname}
\def\numstyle#1{\usename{@\expandafter\blank\string#1style}#1}
\def\ifblank#1\then{\useafter\ifx{@\expandafter\blank\string#1}\blank}

\def\blank#1{}

\def\Roman#1{\expandafter\uppercase\expandafter{\romannumeral#1}}
\def\alphabetic#1{%
     \ifcase#1
          $\bullet$\err@badcountervalue{alphabetic}%
          \or a\or b\or c\or d\or e\or f\or g\or h\or i\or j\or k\or l\or m%
          \or n\or o\or p\or q\or r\or s\or t\or u\or v\or w\or x\or y\or z%
     \else$\bullet$\err@badcountervalue{alphabetic}%
     \fi}
\def\Alphabetic#1{\expandafter\uppercase\expandafter{\alphabetic{#1}}}
\def\symbols#1{%
     \ifcase#1
          $\bullet$\err@badcountervalue{symbols}%
          \or*\or\dag\or\ddag\or\S\or$\|$%
          \or**\or\dag\dag\or\ddag\ddag\or\S\S\or$\|\|$%
     \else$\bullet$\err@badcountervalue{symbols}%
     \fi}

%************** String macros *******************************

\catcode`\^^?=13 \def^^?{\relax}

\def\trimleading#1\to#2{\edef#2{#1}%
     \expandafter\@trimleading\expandafter#2#2^^?^^?}
\def\@trimleading#1#2#3^^?{\ifx#2^^?\def#1{}\else\def#1{#2#3}\fi}

\def\trimtrailing#1\to#2{\edef#2{#1}%
     \expandafter\@trimtrailing\expandafter#2#2^^? ^^?\relax}
\def\@trimtrailing#1#2 ^^?#3{\ifx#3\relax\toks@={}%
     \else\def#1{#2}\toks@={\trimtrailing#1\to#1}\fi
     \the\toks@}

\def\trim#1\to#2{\trimleading#1\to#2\trimtrailing#2\to#2}

\catcode`\^^?=15

%************** List macros *********************************

\long\def\additemL#1\to#2{\toks@={\^^\{#1}}\toks@ii=\expandafter{#2}%
     \xdef#2{\the\toks@\the\toks@ii}}

\long\def\additemR#1\to#2{\toks@={\^^\{#1}}\toks@ii=\expandafter{#2}%
     \xdef#2{\the\toks@ii\the\toks@}}

\def\getitemL#1\to#2{\expandafter\@getitemL#1\hack@#1#2}
\def\@getitemL\^^\#1#2\hack@#3#4{\def#4{#1}\def#3{#2}}

%************************************************************
%*
%*             Font-related macros
%*
%************************************************************
\message{font macros,}

%************** Font set-up *********************************

\newdimen\rp@
\newcount\@@sizeindex \@@sizeindex=0
\newcount\@@factori
\newcount\@@factorii
\newcount\@@factoriii
\newcount\@@factoriv

\countdef\maxfam=18
\newfam\itfam
\newfam\bffam
\newfam\bfsfam
\newfam\bmitfam

\def\@mathfontinit{\count@=4
     \loop\textfont\count@=\nullfont
          \scriptfont\count@=\nullfont
          \scriptscriptfont\count@=\nullfont
          \ifnum\count@<\maxfam\advance\count@ by\@ne
     \repeat}

\def\@fontstyleinit{%
     \def\it{\err@fontnotavailable\it}%
     \def\bf{\err@fontnotavailable\bf}%
     \def\bfs{\err@bfstobf}%
     \def\bmit{\err@fontnotavailable\bmit}%
     \def\sc{\err@fontnotavailable\sc}%
     \def\sl{\err@sltoit}%
     \def\ss{\err@fontnotavailable\ss}%
     \def\tt{\err@fontnotavailable\tt}}

\def\@parameterinit#1{\rm\rp@=.1em \@getscaling{#1}%
     \let\^^\=\@doscaling\scalingskipslist
     \setbox\strutbox=\hbox{\vrule
          height.708\baselineskip depth.292\baselineskip width\z@}}

\def\@getfactor#1#2#3#4{\@@factori=#1 \@@factorii=#2
     \@@factoriii=#3 \@@factoriv=#4}

\def\@getscaling#1{\count@=#1 \advance\count@ by-\@@sizeindex\@@sizeindex=#1
     \ifnum\count@<0
          \let\@mulordiv=\divide
          \let\@divormul=\multiply
          \multiply\count@ by\m@ne
     \else\let\@mulordiv=\multiply
          \let\@divormul=\divide
     \fi
     \edef\@@scratcha{\ifcase\count@                {1}{1}{1}{1}\or
          {1}{7}{23}{3}\or     {2}{5}{3}{1}\or      {9}{89}{13}{1}\or
          {6}{25}{6}{1}\or     {8}{71}{14}{1}\or    {6}{25}{36}{5}\or
          {1}{7}{53}{4}\or     {12}{125}{108}{5}\or {3}{14}{53}{5}\or
          {6}{41}{17}{1}\or    {13}{31}{13}{2}\or   {9}{107}{71}{2}\or
          {11}{139}{124}{3}\or {1}{6}{43}{2}\or     {10}{107}{42}{1}\or
          {1}{5}{43}{2}\or     {5}{69}{65}{1}\or    {11}{97}{91}{2}\fi}%
     \expandafter\@getfactor\@@scratcha}

\def\@doscaling#1{\@mulordiv#1by\@@factori\@divormul#1by\@@factorii
     \@mulordiv#1by\@@factoriii\@divormul#1by\@@factoriv}

%************* Size-changing commands ***********************

\newskip\headskip
\newskip\footskip

\def\typesize=#1pt{\count@=#1 \advance\count@ by-10
     \ifcase\count@
          \@setsizex\or\err@badtypesize\or
          \@setsizexii\or\err@badtypesize\or
          \@setsizexiv
     \else\err@badtypesize
     \fi}

\def\@setsizex{\getixpt
     \def\subsubscriptfonts{\vpt}%
          \def\subsubscriptsize{\vpt\@parameterinit{-8}}%
     \def\subscriptfonts{\viipt}\def\subscriptsize{\viipt\@parameterinit{-4}}%
     \def\footnotefonts{\viiipt}\def\footnotesize{\viiipt\@parameterinit{-2}}%
     \def\smallfonts{\ixpt}\def\smallsize{\ixpt\@parameterinit{-1}}%
     \def\normalfonts{\xpt}\def\normalsize{\xpt\@parameterinit{0}}%
     \def\bigfonts{\xiipt}\def\bigsize{\xiipt\@parameterinit{2}}%
     \def\Bigfonts{\xivpt}\def\Bigsize{\xivpt\@parameterinit{4}}%
     \def\biggfonts{\xviipt}\def\biggsize{\xviipt\@parameterinit{6}}%
     \def\Biggfonts{\xxipt}\def\Biggsize{\xxipt\@parameterinit{8}}%
     \def\tinyfonts{\vpt}\def\tinysize{\vpt\@parameterinit{-8}}%
     \def\HUGEFONTS{\xxvpt}\def\HUGESIZE{\xxvpt\@parameterinit{10}}%
     \normalsize\fixedskipslist}

\def\@setsizexii{\getxipt
     \def\subsubscriptfonts{\vipt}%
          \def\subsubscriptsize{\vipt\@parameterinit{-6}}%
     \def\subscriptfonts{\viiipt}%
          \def\subscriptsize{\viiipt\@parameterinit{-2}}%
     \def\footnotefonts{\xpt}\def\footnotesize{\xpt\@parameterinit{0}}%
     \def\smallfonts{\xipt}\def\smallsize{\xipt\@parameterinit{1}}%
     \def\normalfonts{\xiipt}\def\normalsize{\xiipt\@parameterinit{2}}%
     \def\bigfonts{\xivpt}\def\bigsize{\xivpt\@parameterinit{4}}%
     \def\Bigfonts{\xviipt}\def\Bigsize{\xviipt\@parameterinit{6}}%
     \def\biggfonts{\xxipt}\def\biggsize{\xxipt\@parameterinit{8}}%
     \def\Biggfonts{\xxvpt}\def\Biggsize{\xxvpt\@parameterinit{10}}%
     \def\tinyfonts{\vpt}\def\tinysize{\vpt\@parameterinit{-8}}%
     \def\HUGEFONTS{\xxvpt}\def\HUGESIZE{\xxvpt\@parameterinit{10}}%
     \normalsize\fixedskipslist}

\def\@setsizexiv{\getxiiipt
     \def\subsubscriptfonts{\viipt}%
          \def\subsubscriptsize{\viipt\@parameterinit{-4}}%
     \def\subscriptfonts{\xpt}\def\subscriptsize{\xpt\@parameterinit{0}}%
     \def\footnotefonts{\xiipt}\def\footnotesize{\xiipt\@parameterinit{2}}%
     \def\smallfonts{\xiiipt}\def\smallsize{\xiiipt\@parameterinit{3}}%
     \def\normalfonts{\xivpt}\def\normalsize{\xivpt\@parameterinit{4}}%
     \def\bigfonts{\xviipt}\def\bigsize{\xviipt\@parameterinit{6}}%
     \def\Bigfonts{\xxipt}\def\Bigsize{\xxipt\@parameterinit{8}}%
     \def\biggfonts{\xxvpt}\def\biggsize{\xxvpt\@parameterinit{10}}%
     \def\Biggfonts{\err@sizetoolarge\Biggfonts\HUGEFONTS}%
          \def\Biggsize{\err@sizetoolarge\Biggsize\HUGESIZE}%
     \def\tinyfonts{\vpt}\def\tinysize{\vpt\@parameterinit{-8}}%
     \def\HUGEFONTS{\xxvpt}\def\HUGESIZE{\xxvpt\@parameterinit{10}}%
     \normalsize\fixedskipslist}

\def\subsubscriptfonts{\vpt} \def\subsubscriptsize{\vpt\@parameterinit{-8}}
\def\subscriptfonts{\viipt}  \def\subscriptsize{\viipt\@parameterinit{-4}}
\def\footnotefonts{\viiipt}  \def\footnotesize{\viiipt\@parameterinit{-2}}
\def\smallfonts{\err@sizenotavailable\smallfonts}
                             \def\smallsize{\ixpt\@parameterinit{-1}}
\def\normalfonts{\xpt}       \def\normalsize{\xpt\@parameterinit{0}}
\def\bigfonts{\xiipt}        \def\bigsize{\xiipt\@parameterinit{2}}
\def\Bigfonts{\xivpt}        \def\Bigsize{\xivpt\@parameterinit{4}}
\def\biggfonts{\xviipt}      \def\biggsize{\xviipt\@parameterinit{6}}
\def\Biggfonts{\xxipt}       \def\Biggsize{\xxipt\@parameterinit{8}}
\def\tinyfonts{\vpt}         \def\tinysize{\vpt\@parameterinit{-8}}
\def\HUGEFONTS{\xxvpt}       \def\HUGESIZE{\xxvpt\@parameterinit{10}}

%************************************************************
%*
%*             Document layout
%*
%************************************************************
\message{document layout,}

%************** Page format *********************************

\newtoks\everyoutput \everyoutput={}
\newdimen\depthofpage
\newcount\pagenum \pagenum=0

\newdimen\oddtopmargin  \newdimen\eventopmargin
\newdimen\oddleftmargin \newdimen\evenleftmargin
\newtoks\oddhead        \newtoks\evenhead
\newtoks\oddfoot        \newtoks\evenfoot

\def\topmargin{\afterassignment\@seteventop\oddtopmargin}
\def\leftmargin{\afterassignment\@setevenleft\oddleftmargin}
\def\head{\afterassignment\@setevenhead\oddhead}
\def\foot{\afterassignment\@setevenfoot\oddfoot}

\def\@seteventop{\eventopmargin=\oddtopmargin}
\def\@setevenleft{\evenleftmargin=\oddleftmargin}
\def\@setevenhead{\evenhead=\oddhead}
\def\@setevenfoot{\evenfoot=\oddfoot}

\def\pagenumstyle#1{\@setnumstyle\pagenum{#1}}

\newif\ifdraft
\def\draft{\drafttrue\leftmargin=.5in \overfullrule=5pt }

\def\outputstyle#1{\global\expandafter\let\expandafter
          \@outputstyle\csname#1output\endcsname
     \usename{#1setup}}

\output={\@outputstyle}

\def\normaloutput{\the\everyoutput
     \global\advance\pagenum by\@ne
     \ifodd\pagenum
          \voffset=\oddtopmargin \hoffset=\oddleftmargin
     \else\voffset=\eventopmargin \hoffset=\evenleftmargin
     \fi
     \advance\voffset by-1in  \advance\hoffset by-1in
     \count0=\pagenum
     \expandafter\shipout\pagebox
     \ifnum\outputpenalty>-\@MM\else\dosupereject\fi}

\newdimen\fullhsize
\newbox\leftpage
\newcount\leftpagenum
\newcount\outputpagenum \outputpagenum=0
\let\leftorright=L

\def\twoupoutput{\the\everyoutput
     \global\advance\pagenum by\@ne
     \if L\leftorright
          \global\setbox\leftpage=\leftline{\pagebox}%
          \global\leftpagenum=\pagenum
          \global\let\leftorright=R%
     \else\global\advance\outputpagenum by\@ne
          \ifodd\outputpagenum
               \voffset=\oddtopmargin \hoffset=\oddleftmargin
          \else\voffset=\eventopmargin \hoffset=\evenleftmargin
          \fi
          \advance\voffset by-1in  \advance\hoffset by-1in
          \count0=\leftpagenum \count1=\pagenum
          \shipout\vbox{\hbox to\fullhsize
               {\box\leftpage\hfil\leftline{\pagebox}}}%
          \global\let\leftorright=L%
     \fi
     \ifnum\outputpenalty>-\@MM
     \else\dosupereject
          \if R\leftorright
               \globaldefs=\@ne\head={\hfil}\foot={\hfil}\globaldefs=\z@
               \null\newpage
          \fi
     \fi}

\def\pagebox{\vbox{\makeheadline\pagebody\makefootline}}

\def\makeheadline{%
     \vbox to\z@{\baselinestretch=\@m
          \vskip\topskip\vskip-.708\baselineskip\vskip-\headskip
          \line{\vbox to\ht\strutbox{}%
               \ifodd\pagenum\the\oddhead\else\the\evenhead\fi}%
          \vss}%
     \nointerlineskip}

\def\pagebody{\vbox to\vsize{%
     \boxmaxdepth\maxdepth
     \ifvoid\topins\else\unvbox\topins\fi
     \depthofpage=\dp255
     \unvbox255
     \ifraggedbottom\kern-\depthofpage\vfil\fi
     \ifvoid\footins
     \else\vskip\skip\footins
          \footnoterule
          \unvbox\footins
          \vskip-\footnoteskip
     \fi}}

\def\makefootline{\baselineskip=\footskip
     \line{\ifodd\pagenum\the\oddfoot\else\the\evenfoot\fi}}

%************** Sectioning commands *************************

\newskip\abovechapterskip
\newskip\belowchapterskip
\newskip\abovesectionskip
\newskip\belowsectionskip
\newskip\abovesubsectionskip
\newskip\belowsubsectionskip

\def\chapterstyle#1{\global\expandafter\let\expandafter\@chapterstyle
     \csname#1text\endcsname}
\def\sectionstyle#1{\global\expandafter\let\expandafter\@sectionstyle
     \csname#1text\endcsname}
\def\subsectionstyle#1{\global\expandafter\let\expandafter\@subsectionstyle
     \csname#1text\endcsname}

\def\chapter#1{%
     \ifdim\lastskip=17sp \else\chapterbreak\vskip\abovechapterskip\fi
     \@chapterstyle{\ifblank\chapternumstyle\then
          \else\newchapternum=\next\chapternumformat\ \fi#1}%
     \nobreak\vskip\belowchapterskip\vskip17sp }

\def\section#1{%
     \ifdim\lastskip=17sp \else\sectionbreak\vskip\abovesectionskip\fi
     \@sectionstyle{\ifblank\sectionnumstyle\then
          \else\newsectionnum=\next\sectionnumformat\ \fi#1}%
     \nobreak\vskip\belowsectionskip\vskip17sp }

\def\subsection#1{%
     \ifdim\lastskip=17sp \else\subsectionbreak\vskip\abovesubsectionskip\fi
     \@subsectionstyle{\ifblank\subsectionnumstyle\then
          \else\newsubsectionnum=\next\subsectionnumformat\ \fi#1}%
     \nobreak\vskip\belowsubsectionskip\vskip17sp }

%************** Text formatting commands ********************

\let\TeXunderline=\underline
\let\TeXoverline=\overline
\def\underline#1{\relax\ifmmode\TeXunderline{#1}\else
     $\TeXunderline{\hbox{#1}}$\fi}
\def\overline#1{\relax\ifmmode\TeXoverline{#1}\else
     $\TeXoverline{\hbox{#1}}$\fi}

\def\baselinestretch{\afterassignment\@baselinestretch\count@}
\def\@baselinestretch{\baselineskip=\normalbaselineskip
     \divide\baselineskip by\@m\baselineskip=\count@\baselineskip
     \setbox\strutbox=\hbox{\vrule
          height.708\baselineskip depth.292\baselineskip width\z@}%
     \bigskipamount=\the\baselineskip
          plus.25\baselineskip minus.25\baselineskip
     \medskipamount=.5\baselineskip
          plus.125\baselineskip minus.125\baselineskip
     \smallskipamount=.25\baselineskip
          plus.0625\baselineskip minus.0625\baselineskip}

\def\\{\ifhmode\ifnum\lastpenalty=-\@M\else\hfil\penalty-\@M\fi\fi
     \ignorespaces}
\def\newpage{\vfil\break}

\def\lefttext#1{\par{\@text\leftskip=\z@\rightskip=\centering
     \noindent#1\par}}
\def\righttext#1{\par{\@text\leftskip=\centering\rightskip=\z@
     \noindent#1\par}}
\def\centertext#1{\par{\@text\leftskip=\centering\rightskip=\centering
     \noindent#1\par}}
\def\@text{\parindent=\z@ \parfillskip=\z@ \everypar={}%
     \spaceskip=.3333em \xspaceskip=.5em
     \def\\{\ifhmode\ifnum\lastpenalty=-\@M\else\penalty-\@M\fi\fi
          \ignorespaces}}

\def\beginleft{\par\@text\leftskip=\z@ \rightskip=\centering}
     
\def\beginright{\par\@text\leftskip=\centering\rightskip=\z@ }
     
\def\begincenter{\par\@text\leftskip=\centering\rightskip=\centering}

\def\beginnarrow{\defaultoption[\parindent]\@beginnarrow}
\def\@beginnarrow[#1]{\par\advance\leftskip by#1\advance\rightskip by#1}

\begingroup
\catcode`\[=1 \catcode`\{=11
\gdef\beginignore[\endgroup\bgroup
     \catcode`\e=0 \catcode`\\=12 \catcode`\{=11 \catcode`\f=12 \let\or=\relax
     \let\nd{ignor=\fi \let\}=\egroup
     \iffalse}
\endgroup

\long\def\marginnote#1{\leavevmode
     \edef\@marginsf{\spacefactor=\the\spacefactor\relax}%
     \ifdraft\strut\vadjust{%
          \hbox to\z@{\hskip\hsize\hskip.1in
               \vbox to\z@{\vskip-\dp\strutbox
                    \marginnoteformat
                    \vskip-\ht\strutbox
                    \noindent\strut#1\par
                    \vss}%
               \hss}}%
     \fi
     \@marginsf}

%************** The \bye command ****************************

\newtoks\everybye \everybye={\par\vfil}
\outer\def\bye{\the\everybye
     \footnotecheck
     \prelabelcheck
     \streamcheck
     \supereject
     \TeXend}

%************************************************************
%*
%*             Footnotes
%*
%************************************************************
\message{footnotes,}

\newcount\footnotenum \footnotenum=0
\newskip\footnoteskip
\let\@footnotelist=\empty

\def\footnotenumstyle#1{\@setnumstyle\footnotenum{#1}%
     \useafter\ifx{@footnotenumstyle}\symbols
          \global\let\@footup=\empty
     \else\global\let\@footup=\markup
     \fi}

\def\footnote{\footnotecheck\defaultoption[]\@footnote}
\def\@footnote[#1]{\@footnotemark[#1]\@footnotetext}

\def\footnotemark{\defaultoption[]\@footnotemark}
\def\@footnotemark[#1]{\let\@footsf=\empty
     \ifhmode\edef\@footsf{\spacefactor=\the\spacefactor\relax}\/\fi
     \ifnoarg#1\then
          \global\advance\footnotenum by\@ne
          \@footup{\footnotenumformat}%
          \edef\@@foota{\footnotenum=\the\footnotenum\relax}%
          \expandafter\additemR\expandafter\@footup\expandafter
               {\@@foota\footnotenumformat}\to\@footnotelist
          \global\let\@footnotelist=\@footnotelist
     \else\markup{#1}%
          \additemR\markup{#1}\to\@footnotelist
          \global\let\@footnotelist=\@footnotelist
     \fi
     \@footsf}

\def\footnotetext{%
     \ifx\@footnotelist\empty\err@extrafootnotetext\else\@footnotetext\fi}
\def\@footnotetext{%
     \getitemL\@footnotelist\to\@@foota
     \global\let\@footnotelist=\@footnotelist
     \insert\footins\bgroup
     \footnoteformat
     \splittopskip=\ht\strutbox\splitmaxdepth=\dp\strutbox
     \interlinepenalty=\interfootnotelinepenalty\floatingpenalty=\@MM
     \noindent\llap{\@@foota}\strut
     \bgroup\aftergroup\@footnoteend
     \let\@@scratcha=}
\def\@footnoteend{\strut\par\vskip\footnoteskip\egroup}

\def\footnoterule{\normalfonts
     \kern-.3em \hrule width2in height.04em \kern .26em }

\def\footnotecheck{%
     \ifx\@footnotelist\empty
     \else\err@extrafootnotemark
          \global\let\@footnotelist=\empty
     \fi}

%************************************************************
%*
%*             Labelling macros
%*
%************************************************************
\message{labels,}

\let\@@labeldef=\xdef
\newif\if@labelfile
\newwrite\@labelfile
\let\@prelabellist=\empty

\def\label#1#2{\trim#1\to\@@labarg\edef\@@labtext{#2}%
     \edef\@@labname{lab@\@@labarg}%
     \useafter\ifundefined\@@labname\then\else\@yeslab\fi
     \useafter\@@labeldef\@@labname{#2}%
     \ifstreaming
          \expandafter\toks@\expandafter\expandafter\expandafter
               {\csname\@@labname\endcsname}%
          \immediate\write\streamout{\noexpand\label{\@@labarg}{\the\toks@}}%
     \fi}
\def\@yeslab{%
     \useafter\ifundefined{if\@@labname}\then
          \err@labelredef\@@labarg
     \else\useif{if\@@labname}\then
               \err@labelredef\@@labarg
          \else\global\usename{\@@labname true}%
               \useafter\ifundefined{pre\@@labname}\then
               \else\useafter\ifx{pre\@@labname}\@@labtext
                    \else\err@badlabelmatch\@@labarg
                    \fi
               \fi
               \if@labelfile
               \else\global\@labelfiletrue
                    \immediate\write\sixt@@n{--> Creating file \jobname.lab}%
                    \immediate\openout\@labelfile=\jobname.lab
               \fi
               \immediate\write\@labelfile
                    {\noexpand\prelabel{\@@labarg}{\@@labtext}}%
          \fi
     \fi}

\def\putlab#1{\trim#1\to\@@labarg\edef\@@labname{lab@\@@labarg}%
     \useafter\ifundefined\@@labname\then\@nolab\else\usename\@@labname\fi}
\def\@nolab{%
     \useafter\ifundefined{pre\@@labname}\then
          \undefinedlabelformat
          \err@needlabel\@@labarg
          \useafter\xdef\@@labname{\undefinedlabelformat}%
     \else\usename{pre\@@labname}%
          \useafter\xdef\@@labname{\usename{pre\@@labname}}%
     \fi
     \useafter\newif{if\@@labname}%
     \expandafter\additemR\@@labarg\to\@prelabellist}

\def\prelabel#1{\useafter\gdef{prelab@#1}}

\def\ifundefinedlabel#1\then{%
     \expandafter\ifx\csname lab@#1\endcsname\relax}
\def\useiflab#1\then{\csname iflab@#1\endcsname}

\def\prelabelcheck{{%
     \def\^^\##1{\useiflab{##1}\then\else\err@undefinedlabel{##1}\fi}%
     \@prelabellist}}

%************************************************************
%*
%*             Equation numbering
%*
%************************************************************
\message{equation numbering,}

\newcount\chapternum
\newcount\sectionnum
\newcount\subsectionnum
\newcount\equationnum
\newcount\subequationnum
\newcount\figurenum
\newcount\subfigurenum
\newcount\tablenum
\newcount\subtablenum

\newif\if@subeqncount
\newif\if@subfigcount
\newif\if@subtblcount

\def\newchapternum{\newsectionnum=\z@\@resetnum\chapternum}
\def\newsectionnum{\newsubsectionnum=\z@\@resetnum\sectionnum}
\def\newsubsectionnum{\newequationnum=\z@\newfigurenum=\z@\newtablenum=\z@
     \@resetnum\subsectionnum}
\def\newequationnum{\newsubequationnum=\z@\@resetnum\equationnum}
\def\newsubequationnum{\@resetnum\subequationnum}
\def\newfigurenum{\newsubfigurenum=\z@\@resetnum\figurenum}
\def\newsubfigurenum{\@resetnum\subfigurenum}
\def\newtablenum{\newsubtablenum=\z@\@resetnum\tablenum}
\def\newsubtablenum{\@resetnum\subtablenum}

\def\@resetnum#1{\global\advance#1by1 \edef\next{\the#1\relax}\global#1}

\newchapternum=0

\def\chapternumstyle#1{\@setnumstyle\chapternum{#1}}
\def\sectionnumstyle#1{\@setnumstyle\sectionnum{#1}}
\def\subsectionnumstyle#1{\@setnumstyle\subsectionnum{#1}}
\def\equationnumstyle#1{\@setnumstyle\equationnum{#1}}
\def\subequationnumstyle#1{\@setnumstyle\subequationnum{#1}%
     \ifblank\subequationnumstyle\then\global\@subeqncountfalse\fi
     \ignorespaces}
\def\figurenumstyle#1{\@setnumstyle\figurenum{#1}}
\def\subfigurenumstyle#1{\@setnumstyle\subfigurenum{#1}%
     \ifblank\subfigurenumstyle\then\global\@subfigcountfalse\fi
     \ignorespaces}
\def\tablenumstyle#1{\@setnumstyle\tablenum{#1}}
\def\subtablenumstyle#1{\@setnumstyle\subtablenum{#1}%
     \ifblank\subtablenumstyle\then\global\@subtblcountfalse\fi
     \ignorespaces}

\def\eqnlabel#1{%
     \if@subeqncount
          \newsubequationnum=\next
     \else\newequationnum=\next
          \ifblank\subequationnumstyle\then
          \else\global\@subeqncounttrue
               \newsubequationnum=\@ne
          \fi
     \fi
     \label{#1}{\puteqnformat}(\puteqn{#1})%
     \ifdraft\rlap{\hskip.1in{\tt#1}}\fi}

\let\puteqn=\putlab

\def\equation#1#2{\useafter\gdef{eqn@#1}{#2\eqno\eqnlabel{#1}}}
\def\Equation#1{\useafter\gdef{eqn@#1}}

\def\putequation#1{\useafter\ifundefined{eqn@#1}\then
     \err@undefinedeqn{#1}\else\usename{eqn@#1}\fi}

\def\eqnseriesstyle#1{\gdef\@eqnseriesstyle{#1}}
\def\begineqnseries{\subequationnumstyle{\@eqnseriesstyle}%
     \defaultoption[]\@begineqnseries}
\def\@begineqnseries[#1]{\edef\@@eqnname{#1}}
\def\endeqnseries{\subequationnumstyle{blank}%
     \expandafter\ifnoarg\@@eqnname\then
     \else\label\@@eqnname{\puteqnformat}%
     \fi
     \aftergroup\ignorespaces}

\def\figlabel#1{%
     \if@subfigcount
          \newsubfigurenum=\next
     \else\newfigurenum=\next
          \ifblank\subfigurenumstyle\then
          \else\global\@subfigcounttrue
               \newsubfigurenum=\@ne
          \fi
     \fi
     \label{#1}{\putfigformat}\putfig{#1}%
     {\def\marginnoteformat{\tt}\marginnote{#1}}}

\let\putfig=\putlab

\def\figseriesstyle#1{\gdef\@figseriesstyle{#1}}
\def\beginfigseries{\subfigurenumstyle{\@figseriesstyle}%
     \defaultoption[]\@beginfigseries}
\def\@beginfigseries[#1]{\edef\@@figname{#1}}
\def\endfigseries{\subfigurenumstyle{blank}%
     \expandafter\ifnoarg\@@figname\then
     \else\label\@@figname{\putfigformat}%
     \fi
     \aftergroup\ignorespaces}

\def\tbllabel#1{%
     \if@subtblcount
          \newsubtablenum=\next
     \else\newtablenum=\next
          \ifblank\subtablenumstyle\then
          \else\global\@subtblcounttrue
               \newsubtablenum=\@ne
          \fi
     \fi
     \label{#1}{\puttblformat}\puttbl{#1}%
     {\def\marginnoteformat{\tt}\marginnote{#1}}}

\let\puttbl=\putlab

\def\tblseriesstyle#1{\gdef\@tblseriesstyle{#1}}
\def\begintblseries{\subtablenumstyle{\@tblseriesstyle}%
     \defaultoption[]\@begintblseries}
\def\@begintblseries[#1]{\edef\@@tblname{#1}}
\def\endtblseries{\subtablenumstyle{blank}%
     \expandafter\ifnoarg\@@tblname\then
     \else\label\@@tblname{\puttblformat}%
     \fi
     \aftergroup\ignorespaces}

%************************************************************
%*
%*             Reference numbering
%*
%************************************************************
\message{reference numbering,}

\newcount\referencenum \referencenum=0
\newcount\@@prerefcount \@@prerefcount=0
\newcount\@@thisref
\newcount\@@lastref
\newcount\@@loopref
\newcount\@@refseq
\newdimen\refnumindent
\let\@undefreflist=\empty

\def\referencenumstyle#1{\@setnumstyle\referencenum{#1}}

\def\referencestyle#1{\usename{@ref#1}}

\def\@refsequential{%
     \gdef\@refpredef##1{\global\advance\referencenum by\@ne
          \let\^^\=0\label{##1}{\^^\{\the\referencenum}}%
          \useafter\gdef{ref@\the\referencenum}{{##1}{\undefinedlabelformat}}}%
     \gdef\@reference##1##2{%
          \ifundefinedlabel##1\then
          \else\def\^^\####1{\global\@@thisref=####1\relax}\putlab{##1}%
               \useafter\gdef{ref@\the\@@thisref}{{##1}{##2}}%
          \fi}%
     \gdef\endputreferences{%
          \loop\ifnum\@@loopref<\referencenum
                    \advance\@@loopref by\@ne
                    \expandafter\expandafter\expandafter\@printreference
                         \csname ref@\the\@@loopref\endcsname
          \repeat
          \par}}

\def\@refpreordered{%
     \gdef\@refpredef##1{\global\advance\referencenum by\@ne
          \additemR##1\to\@undefreflist}%
     \gdef\@reference##1##2{%
          \ifundefinedlabel##1\then
          \else\global\advance\@@loopref by\@ne
               {\let\^^\=0\label{##1}{\^^\{\the\@@loopref}}}%
               \@printreference{##1}{##2}%
          \fi}
     \gdef\endputreferences{%
          \def\^^\####1{\useiflab{####1}\then
               \else\reference{####1}{\undefinedlabelformat}\fi}%
          \@undefreflist
          \par}}

\def\beginprereferences{\par
     \def\reference##1##2{\global\advance\referencenum by1\@ne
          \let\^^\=0\label{##1}{\^^\{\the\referencenum}}%
          \useafter\gdef{ref@\the\referencenum}{{##1}{##2}}}}
\def\endprereferences{\global\@@prerefcount=\the\referencenum\par}

\def\beginputreferences{\par
     \refnumindent=\z@\@@loopref=\z@
     \loop\ifnum\@@loopref<\referencenum
               \advance\@@loopref by\@ne
               \setbox\z@=\hbox{\referencenum=\@@loopref
                    \referencenumformat\enskip}%
               \ifdim\wd\z@>\refnumindent\refnumindent=\wd\z@\fi
     \repeat
     \putreferenceformat
     \@@loopref=\z@
     \loop\ifnum\@@loopref<\@@prerefcount
               \advance\@@loopref by\@ne
               \expandafter\expandafter\expandafter\@printreference
                    \csname ref@\the\@@loopref\endcsname
     \repeat
     \let\reference=\@reference}

\def\@printreference#1#2{\ifx#2\undefinedlabelformat\err@undefinedref{#1}\fi
     \noindent\ifdraft\rlap{\hskip\hsize\hskip.1in \tt#1}\fi
     \llap{\referencenum=\@@loopref\referencenumformat\enskip}#2\par}

\def\reference#1#2{{\par\refnumindent=\z@\putreferenceformat\noindent#2\par}}

\def\putref#1{\trim#1\to\@@refarg
     \expandafter\ifnoarg\@@refarg\then
          \toks@={\relax}%
     \else\@@lastref=-\@m\def\@@refsep{}\def\@more{\@nextref}%
          \toks@={\@nextref#1,,}%
     \fi\the\toks@}
\def\@nextref#1,{\trim#1\to\@@refarg
     \expandafter\ifnoarg\@@refarg\then
          \let\@more=\relax
     \else\ifundefinedlabel\@@refarg\then
               \expandafter\@refpredef\expandafter{\@@refarg}%
          \fi
          \def\^^\##1{\global\@@thisref=##1\relax}%
          \global\@@thisref=\m@ne
          \setbox\z@=\hbox{\putlab\@@refarg}%
     \fi
     \advance\@@lastref by\@ne
     \ifnum\@@lastref=\@@thisref\advance\@@refseq by\@ne\else\@@refseq=\@ne\fi
     \ifnum\@@lastref<\z@
     \else\ifnum\@@refseq<\thr@@
               \@@refsep\def\@@refsep{,}%
               \ifnum\@@lastref>\z@
                    \advance\@@lastref by\m@ne
                    {\referencenum=\@@lastref\putrefformat}%
               \else\undefinedlabelformat
               \fi
          \else\def\@@refsep{--}%
          \fi
     \fi
     \@@lastref=\@@thisref
     \@more}

%************************************************************
%*
%*             Job streaming
%*
%************************************************************
\message{streaming,}

\newif\ifstreaming

\def\streamto{\defaultoption[\jobname]\@streamto}
\def\@streamto[#1]{\global\streamingtrue
     \immediate\write\sixt@@n{--> Streaming to #1.str}%
     \newwrite\streamout\immediate\openout\streamout=#1.str }

\def\streamfrom{\defaultoption[\jobname]\@streamfrom}
\def\@streamfrom[#1]{\newread\streamin\openin\streamin=#1.str
     \ifeof\streamin
          \expandafter\err@nostream\expandafter{#1.str}%
     \else\immediate\write\sixt@@n{--> Streaming from #1.str}%
          \let\@@labeldef=\gdef
          \ifstreaming
               \edef\@elc{\endlinechar=\the\endlinechar}%
               \endlinechar=\m@ne
               \loop\read\streamin to\@@scratcha
                    \ifeof\streamin
                         \streamingfalse
                    \else\toks@=\expandafter{\@@scratcha}%
                         \immediate\write\streamout{\the\toks@}%
                    \fi
                    \ifstreaming
               \repeat
               \@elc
               \input #1.str
               \streamingtrue
          \else\input #1.str
          \fi
          \let\@@labeldef=\xdef
     \fi}

\def\streamcheck{\ifstreaming
     \immediate\write\streamout{\pagenum=\the\pagenum}%
     \immediate\write\streamout{\footnotenum=\the\footnotenum}%
     \immediate\write\streamout{\referencenum=\the\referencenum}%
     \immediate\write\streamout{\chapternum=\the\chapternum}%
     \immediate\write\streamout{\sectionnum=\the\sectionnum}%
     \immediate\write\streamout{\subsectionnum=\the\subsectionnum}%
     \immediate\write\streamout{\equationnum=\the\equationnum}%
     \immediate\write\streamout{\subequationnum=\the\subequationnum}%
     \immediate\write\streamout{\figurenum=\the\figurenum}%
     \immediate\write\streamout{\subfigurenum=\the\subfigurenum}%
     \immediate\write\streamout{\tablenum=\the\tablenum}%
     \immediate\write\streamout{\subtablenum=\the\subtablenum}%
     \immediate\closeout\streamout
     \fi}

%************************************************************
%*
%*             Error messages
%*
%************************************************************

\def\err@badtypesize{%
     \errhelp={The limited availability of certain fonts requires^^J%
          that the base type size be 10pt, 12pt, or 14pt.^^J}%
     \errmessage{--> Illegal base type size}}

\def\err@badsizechange{\immediate\write\sixt@@n
     {--> Size change not allowed in math mode, ignored}}

\def\err@sizetoolarge#1{\immediate\write\sixt@@n
     {--> \noexpand#1 too big, substituting HUGE}}

\def\err@sizenotavailable#1{\immediate\write\sixt@@n
     {--> Size not available, \noexpand#1 ignored}}

\def\err@fontnotavailable#1{\immediate\write\sixt@@n
     {--> Font not available, \noexpand#1 ignored}}

\def\err@sltoit{\immediate\write\sixt@@n
     {--> Style \noexpand\sl not available, substituting \noexpand\it}%
     \it}

\def\err@bfstobf{\immediate\write\sixt@@n
     {--> Style \noexpand\bfs not available, substituting \noexpand\bf}%
     \bf}

\def\err@badgroup#1#2{%
     \errhelp={The block you have just tried to close was not the one^^J%
          most recently opened.^^J}%
     \errmessage{--> \noexpand\end{#1} doesn't match \noexpand\begin{#2}}}

\def\err@badcountervalue#1{\immediate\write\sixt@@n
     {--> Counter (#1) out of bounds}}

\def\err@extrafootnotemark{\immediate\write\sixt@@n
     {--> \noexpand\footnotemark command
          has no corresponding \noexpand\footnotetext}}

\def\err@extrafootnotetext{%
     \errhelp{You have given a \noexpand\footnotetext command without first
          specifying^^Ja \noexpand\footnotemark.^^J}%
     \errmessage{--> \noexpand\footnotetext command has no corresponding
          \noexpand\footnotemark}}

\def\err@labelredef#1{\immediate\write\sixt@@n
     {--> Label "#1" redefined}}

\def\err@badlabelmatch#1{\immediate\write\sixt@@n
     {--> Definition of label "#1" doesn't match value in \jobname.lab}}

\def\err@needlabel#1{\immediate\write\sixt@@n
     {--> Label "#1" cited before its definition}}

\def\err@undefinedlabel#1{\immediate\write\sixt@@n
     {--> Label "#1" cited but never defined}}

\def\err@undefinedeqn#1{\immediate\write\sixt@@n
     {--> Equation "#1" not defined}}

\def\err@undefinedref#1{\immediate\write\sixt@@n
     {--> Reference "#1" not defined}}

\def\err@nostream#1{%
     \errhelp={You have tried to input a stream file that doesn't exist.^^J}%
     \errmessage{--> Stream file #1 not found}}

%************************************************************
%*
%*             Initialization
%*
%************************************************************
\message{jyTeX initialization}

\everyjob{\immediate\write16{--> jyTeX version \fmtversion}%
     \edef\@@jobname{\jobname}%
%     \openin0=\inputpath jysupp
%     \ifeof0
%     \else\closein0
%          \immediate\write16{--> Additional macros loaded from jysupp.tex}%
%          \jyinput jysupp
%     \fi
%     \openin0=\inputpath jylocal
%     \ifeof0
%     \else\closein0
%          \immediate\write16{--> Additional macros loaded from jylocal.tex}%
%          \jyinput jylocal
%     \fi
     \edef\jobname{\@@jobname}%
     \settime
     \openin0=\jobname.lab
     \ifeof0
     \else\closein0
          \immediate\write16{--> Getting labels from file \jobname.lab}%
          \input\jobname.lab
     \fi}

%************** Spacing *************************************

\def\fixedskipslist{%
     \^^\{\topskip}%
     \^^\{\splittopskip}%
     \^^\{\maxdepth}%
     \^^\{\skip\topins}%
     \^^\{\skip\footins}%
     \^^\{\headskip}%
     \^^\{\footskip}}

\def\scalingskipslist{%
     \^^\{\p@renwd}%
     \^^\{\delimitershortfall}%
     \^^\{\nulldelimiterspace}%
     \^^\{\scriptspace}%
     \^^\{\jot}%
     \^^\{\normalbaselineskip}%
     \^^\{\normallineskip}%
     \^^\{\normallineskiplimit}%
     \^^\{\baselineskip}%
     \^^\{\lineskip}%
     \^^\{\lineskiplimit}%
     \^^\{\bigskipamount}%
     \^^\{\medskipamount}%
     \^^\{\smallskipamount}%
     \^^\{\parskip}%
     \^^\{\parindent}%
     \^^\{\abovedisplayskip}%
     \^^\{\belowdisplayskip}%
     \^^\{\abovedisplayshortskip}%
     \^^\{\belowdisplayshortskip}%
     \^^\{\abovechapterskip}%
     \^^\{\belowchapterskip}%
     \^^\{\abovesectionskip}%
     \^^\{\belowsectionskip}%
     \^^\{\abovesubsectionskip}%
     \^^\{\belowsubsectionskip}}

%************** Document layout *****************************

\def\twoupsetup{%                                % setup for twoup style
     \topmargin=.75in
     \leftmargin=.5in
     \vsize=6.9in
     \hsize=4.75in
     \fullhsize=10in
     \let\draft=\relax}

\outputstyle{normal}                             % page style

\def\marginnoteformat{\subscriptsize             % paragraphing of margin notes
     \hsize=1in \baselinestretch=1000 \everypar={}%
     \tolerance=5000 \hbadness=5000 \parskip=0pt \parindent=0pt
     \leftskip=0pt \rightskip=0pt \raggedright}

\head={\ifdraft\normalfonts\it\hfil DRAFT\hfil   % format of headline
     \llap{\number\day\ \monthword\month\ \militarytime}\else\hfil\fi}
\foot={\hfil\normalfonts\numstyle\pagenum\hfil}  % format of footline

\normalbaselineskip=12pt                         % usual \baselineskip
\normallineskip=0pt                              % usual \lineskip
\normallineskiplimit=0pt                         % usual \lineskiplimit
\normalbaselines                                 % set \baselineskip

\topskip=.85\baselineskip
\splittopskip=\topskip
\headskip=2\baselineskip
\footskip=\headskip

\pagenumstyle{arabic}                            % counter style

\parskip=0pt                                     % no skip between paragraphs
\parindent=20pt                                  % usual \parindent

\baselinestretch=1000                            % set \big-, \med-, \smallskip

%************** Sectioning **********************************

\chapterstyle{left}                              % position of heading
\chapternumstyle{blank}                          % counter style
\def\chapterbreak{\newpage}                      % break before heading
\abovechapterskip=0pt                            % space before heading
\belowchapterskip=1.5\baselineskip               % space after heading
     plus.38\baselineskip minus.38\baselineskip
\def\chapternumformat{\numstyle\chapternum.}     % format of heading counter

\sectionstyle{left}                              % position of heading
\sectionnumstyle{blank}                          % counter style
\def\sectionbreak{\vskip0pt plus4\baselineskip\penalty-100
     \vskip0pt plus-4\baselineskip}              % break before heading
\abovesectionskip=1.5\baselineskip               % space before heading
     plus.38\baselineskip minus.38\baselineskip
\belowsectionskip=\the\baselineskip              % space after heading
     plus.25\baselineskip minus.25\baselineskip
\def\sectionnumformat{%                          % format of heading counter
     \ifblank\chapternumstyle\then\else\numstyle\chapternum.\fi
     \numstyle\sectionnum.}

\subsectionstyle{left}                           % position of heading
\subsectionnumstyle{blank}                       % counter style
\def\subsectionbreak{\vskip0pt plus4\baselineskip\penalty-100
     \vskip0pt plus-4\baselineskip}              % break before heading
\abovesubsectionskip=\the\baselineskip           % space before heading
     plus.25\baselineskip minus.25\baselineskip
\belowsubsectionskip=.75\baselineskip            % space after heading
     plus.19\baselineskip minus.19\baselineskip
\def\subsectionnumformat{%                       % format of heading counter
     \ifblank\chapternumstyle\then\else\numstyle\chapternum.\fi
     \ifblank\sectionnumstyle\then\else\numstyle\sectionnum.\fi
     \numstyle\subsectionnum.}

%************** Footnotes ***********************************

\footnotenumstyle{symbol}                       % counter style
\footnoteskip=0pt                                % jyTeX spacing parameter
\def\footnotenumformat{\numstyle\footnotenum}    % \footnotemark format
\def\footnoteformat{\footnotesize                % paragraphing of text
     \everypar={}\parskip=0pt \parfillskip=0pt plus1fil
     \leftskip=1em \rightskip=0pt
     \spaceskip=0pt \xspaceskip=0pt
     \def\\{\ifhmode\ifnum\lastpenalty=-10000
          \else\hfil\penalty-10000 \fi\fi\ignorespaces}}

%************** Labels **************************************

\def\undefinedlabelformat{$\bullet$}             % mark for undefined label

%************** Equation numbering **************************

\equationnumstyle{arabic}                        % counter style
\subequationnumstyle{blank}                      % counter style
\figurenumstyle{arabic}                          % counter style
\subfigurenumstyle{blank}                        % counter style
\tablenumstyle{arabic}                           % counter style
\subtablenumstyle{blank}                         % counter style

\eqnseriesstyle{alphabetic}                      % sub-counter style for series
\figseriesstyle{alphabetic}                      % sub-counter style for series
\tblseriesstyle{alphabetic}                      % sub-counter style for series

\def\puteqnformat{\hbox{%                        % equation number format
     \ifblank\chapternumstyle\then\else\numstyle\chapternum.\fi
     \ifblank\sectionnumstyle\then\else\numstyle\sectionnum.\fi
     \ifblank\subsectionnumstyle\then\else\numstyle\subsectionnum.\fi
     \numstyle\equationnum
     \numstyle\subequationnum}}
\def\putfigformat{\hbox{%                        % figure number format
     \ifblank\chapternumstyle\then\else\numstyle\chapternum.\fi
     \ifblank\sectionnumstyle\then\else\numstyle\sectionnum.\fi
     \ifblank\subsectionnumstyle\then\else\numstyle\subsectionnum.\fi
     \numstyle\figurenum
     \numstyle\subfigurenum}}
\def\puttblformat{\hbox{%                        % table number format
     \ifblank\chapternumstyle\then\else\numstyle\chapternum.\fi
     \ifblank\sectionnumstyle\then\else\numstyle\sectionnum.\fi
     \ifblank\subsectionnumstyle\then\else\numstyle\subsectionnum.\fi
     \numstyle\tablenum
     \numstyle\subtablenum}}

%************** Reference numbering *************************

\referencestyle{sequential}                      % referencing method
\referencenumstyle{arabic}                       % counter style
\def\putrefformat{\numstyle\referencenum}        % format of reference citation
\def\referencenumformat{\numstyle\referencenum.} % format of number in list
\def\putreferenceformat{%                        % paragraphing of list
     \everypar={\hangindent=1em \hangafter=1 }%
     \def\\{\hfil\break\null\hskip-1em \ignorespaces}%
     \leftskip=\refnumindent\parindent=0pt \interlinepenalty=1000 }

%************** Font initialization *************************

\normalsize

%*****************************************************************************

\def\fmtversion{2.6M (June 1992)}

\catcode`\@=12
% ------------------ End of jytex.tex -----------------

\input epsf
%\draft
\footnotenumstyle{arabic}
\typesize=10pt
\magnification=1200
\baselineskip=17truept
\hsize=6truein\vsize=8.5truein
%\leftmargin=1.25in
%\oddleftmargin=.5in
%\evenleftmargin=1.5in
\sectionnumstyle{blank}
\chapternumstyle{blank}
\chapternum=1
\sectionnum=1
\pagenum=0
% title style follows

\def\begintitle{\pagenumstyle{blank}\parindent=0pt\begin{narrow}[0.4in]}
\def\endtitle{\end{narrow}\newpage\pagenumstyle{arabic}}

% exercise style follows

\def\beginexercise{\vskip 20truept\parindent=0pt\begin{narrow}[10
truept]}
\def\endexercise{\vskip 10truept\end{narrow}}

% **************    my jyTeX abbreviations   *****************

\def\eql#1{\eqno\eqnlabel{#1}}
\def\ref{\reference}
\def\peq{\puteqn}
\def\pref{\putref}

\def\mgn{\marginnote}
\def\bex{\begin{exercise}}
\def\eex{\end{exercise}}
% *********************** My definitions ************************

\font\open=msbm10 %scaled\magstep1 % For VAX. Borde p195.
 %scaled\magstep1 % For VAX. Borde p195.
%\font\open=msym10 %scaled\magstep1 % For Arbortxt on PC
%\font\opens=msym8 %scaled\magstep1 % For Arbortxt on PC
\font\goth=eufm10  % For Arbortxt on PC and VAX. Borde p199

\def\mbox#1{{\leavevmode\hbox{#1}}}
\def\hspace#1{{\phantom{\mbox#1}}}
\def\oZ{\mbox{\open\char90}}

\def\gC{\mbox{{\goth\char67}}}
\def\gB{\mbox{{\goth\char66}}}

\def\gf{\mbox{{\goth\char102}}}

\def\al{\alpha}
 %in jyTeX
 %in jyTeX
 %in jyTeX
 %in jyTeX
% in jyTeX
\def\be{\beta}
\def\ga{\gamma}

\def\Ga{\Gamma}

\def\ka{\kappa}
\def\la{\lambda}
\def\La{\Lambda}
\def\om{\omega}

\def\th{\theta}

\def\ze{\zeta}

\def\det{{\rm det\,}}

\def\zf{$\zeta$--function}
\def\zfs{$\zeta$--functions}

     % Newline

%\def\frac#1/#2{\leavevmode\kern.1em
%\raise.5ex\hbox{\the\scriptfont0 #1}\kern-.1em/\kern-.15em
%\lower.25ex\hbox{\the\scriptfont0 #2}}
\def\sfrac#1/#2{\leavevmode\kern.1em
\raise.5ex\hbox{\the\scriptscriptfont0 #1}\kern-.1em/\kern-.15em
\lower.25ex\hbox{\the\scriptscriptfont0 #2}}

\def\gtorder{\mathrel{\raise.3ex\hbox{$>$}\mkern-14mu
             \lower0.6ex\hbox{$\sim$}}}
\def\ltorder{\mathrel{\raise.3ex\hbox{$<$}|mkern-14mu
             \lower0.6ex\hbox{\sim$}}}

\def\semidirprod{\rlap{\ss C}\raise1pt\hbox{$\mkern.75mu\times$}}
\def\for{\lower6pt\hbox{$\Big|$}}
\def\fish{\kern-.25em{\phantom{abcde}\over \phantom{abcde}}\kern-.25em}

 %triple dot
 %double dot
 %double dot
%for small #1

\def\boxit#1{\vbox{\hrule\hbox{\vrule\kern3pt
        \vbox{\kern3pt#1\kern3pt}\kern3pt\vrule}\hrule}}
\def\dalemb#1#2{{\vbox{\hrule height .#2pt
        \hbox{\vrule width.#2pt height#1pt \kern#1pt
                \vrule width.#2pt}
        \hrule height.#2pt}}}

\def\ol{\overline}

        %double stroke
 %lower covariant deriv.
    %lower ordinary  deriv.

\def\noin{\noindent}

      %Connection
    %Connection'

\def\cosech{{\rm cosech\,}}
\def\sech{{\rm sech\,}}

\def\eg{{\it e.g. }}
\def\ie{{\it i.e. }}
\def\cf{{\it cf }}
\def\pa{\partial}

 %gives average <#1>
 %gives thermal average <<#1>>
   %gives bracket <#1|#2>
 %gives big bracket <#1|#2>
  %gives
%matrix element <#1|#2|#3>

  %

\def\3j#1#2#3#4#5#6{\left\lgroup\matrix{#1&#2&#3\cr#4&#5&#6\cr}
\right\rgroup}

\def\man{{\cal M}}

\def\caS{{\cal S}}

\def\caD{{\cal D}}

\def\m?{\mgn{?}}

%  *******************  Journal refs **********************

\def\aop#1#2#3{{\it Ann. Phys.} {\bf {#1}} ({#2}) #3}

\def\cmp#1#2#3{{\it Comm. Math. Phys.} {\bf {#1}} ({#2}) #3}
\def\cqg#1#2#3{{\it Class. Quant. Grav.} {\bf {#1}} (19{#2}) #3}

\def\jmp#1#2#3{{\it J. Math. Phys.} {\bf {#1}} (19{#2}) #3}
\def\jpa#1#2#3{{\it J. Phys.} {\bf A{#1}} ({#2}) #3}

\def\np#1#2#3{{\it Nucl. Phys.} {\bf B{#1}} (19{#2}) #3}
\def\pl#1#2#3{{\it Phys. Lett.} {\bf {#1}} (19{#2}) #3}

\def\pr#1#2#3{{\it Phys. Rev.} {\bf {#1}} (19{#2}) #3}
\def\prA#1#2#3{{\it Phys. Rev.} {\bf A{#1}} (19{#2}) #3}

\def\prD#1#2#3{{\it Phys. Rev.} {\bf D{#1}} ({#2}) #3}

\def\prs#1#2#3{{\it Proc. Roy. Soc.} {\bf A{#1}} (19{#2}) #3}
\def\pcps#1#2#3{{\it Proc. Camb. Phil. Soc.} {\bf{#1}} (19{#2}) #3}

\def\dmj#1#2#3{{\it Duke Math. J.} {\bf {#1}} (19{#2}) #3}

\def\jdg#1#2#3{{\it J. Diff. Geom.} {\bf {#1}} (19{#2}) #3}
\def\jfa#1#2#3{{\it J. Func. Anal.} {\bf {#1}} (19{#2}) #3}

\def\jram#1#2#3{{\it J. f. reine u. Angew. Math.} {\bf {#1}} ({#2}) #3}
\def\ma#1#2#3{{\it Math. Ann.} {\bf {#1}} ({#2}) #3}

\def\pams#1#2#3{{\it Proc. Am. Math. Soc.} {\bf {#1}} (19{#2}) #3}

\def\plms#1#2#3{{\it Proc. Lond. Math. Soc.} {\bf {#1}} ({#2}) #3}

\def\tams#1#2#3{{\it Trans. Am. Math. Soc.} {\bf {#1}} (19{#2}) #3}

% *******************   Main text *********************
%\begin{ignore}
\begin{title}
\vglue 20truept
\vskip 100truept
\centertext {\Bigfonts \bf Effective actions on finite cylinders}
\vskip 15truept
\centertext{J.S.Dowker\footnote{ dowkeruk@yahoo.co.uk} }
\vskip 7truept
\centertext{\it Department of Physics and Astronomy \footnote{ Emeritus Professor of Theoretical Physics.},\\
The University of Manchester, Manchester, England.}
\vskip 30truept
%\centertext {Abstract}
\vskip 30truept
\begin{narrow}
Some free--field spectral problems on a generalised cylinder are revisited.

In two dimensions, conformal scalar effective actions for various boundary conditions are written in elliptic function terms and some special values given.

Fermions are then discussed in arbitrary dimensions and an analysis of the most general local (`bag')  boundary conditions is given leading to an intrinsic formula for the eigenvalues which interpolate between Neveu--Schwarz and Ramond. This is used to give reasonably explicit expressions for the effective action and other values of the zeta function. It is shown that these boundary conditions are transferred as a chemical potential to the boundary spectral problem.

The existence of real exponential eigenmodes along the cylinder is pointed out and a curious asymmetry between the positive and negative Dirac spectra uncovered for large cylinder length.
\end{narrow}
\vskip 5truept
\vskip 75truept
\vfil
\end{title}
\pagenum=0
\section{\bf 1. Introduction}\mgn{Rewrite}
The manifold $\gC=I\times\man$ where $I$ is an interval  occurs in a number of different areas of physics and mathematics. If $I$ extends to the whole real line and the metric is Lorentzian, it represents a generalised Einstein Universe. Historically,  $\man$ is then the three--sphere. For a Euclidean signature, it could be called a generalised cylinder and when $\man$ is the circle, S$^1$, it appears, fundamentally, in string theory and in statistical physics where $I$ is often joined up to make a thermal circle so that $\gC$ becomes a torus. For conformal theories, modular invariance then provides a powerful constraint. Cylinders also play an essential  role in the theory of the APS $\eta$--invariant and in the glueing approach to spectral quantities.

Although these topics have been thoroughly explored and re--worked from many points of view,  some very sophisticated, I wish here to present a few observations and connections which, if not unknown, are not always said. My calculations are for the simplest conventional free field theories (scalar and spinor) treated in a very basic way.

The first three sections are concerned with the two--dimensional case and the various elliptical rewritings of the spectral quantities, done mostly for fun. The remaining sections contain somewhat detailed computations of Dirac theory on the finite cylinder, formally in any dimension. The simplest local boundary conditions are treated first and then extended to the most general bag-like ones, for which the functional determinant is calculated as explicitly as possible. The final result is rather simple and may be novel.

\section {\bf2. The logdet in two dimensions}
This is the case on which most attention has been focused and so there is little new to say. However, although it is standard, in one form or another, I write out the conformal scalar effective action on the elementary cylinder, $I\times$S$^1$, for Dirichlet and Neumann conditions,
  $$
  W^{D,N}_{I\times S^1}=-{L\over12r}-\sum_{m=1}^\infty{1\over m(e^{2mL/r}-1)}
  +{1\over2}\log2L\mp{1\over2}\log2\pi r\,.
  \eql{scylinder}
  $$
The length of the cylinder is $L$ and its radius (of the circle cross--section) is $r$. In general terms, the effective action (at one loop) is defined as $W=-\ze_\gC'(0)/2$ where $\ze_\gC$ is the \zf\ on the cylinder..

Instead of the individual D and N expressions, it is sometimes more illuminating to consider their sum and their difference separately because these have distinct geometric significances.

Adding the D and N values gives the expression for the torus, $2L$ now being a circumference,
  $$
  W_{S^1\times S^1}=-{\be\over12r}-2\sum_{m=1}^\infty{1\over m(e^{m\be/r}-1)}
  +\log\be\,,
  \eql{storus}
  $$
where $\be=2L$ can be interpreted as an inverse temperature and the whole expression as a finite temperature thermodynamic free energy (times $\be$) on a spatial circle.

Subtracting $D$ from $N$ gives $\log2\pi r$, which is twice the effective action on the cross--sectional circle of the cylinder. It can be associated with the cylinder {\it boundary} of two circles, (see below).

It is possible, without extra work, to find the effective action for a mixed boundary condition \ie one with N at one end of the cylinder and D at the other. Denoting, for short, cylinder Dirichlet spectral quantities by $(D,D)_L$, Neumann by $(N,N)_L$ and mixed by $(N,D)_L$, the modes are related schematically, by, [\pref{dowhybrid}],
  $$\eqalign{
     (N,D)_L+(D,D)_L&=(D,D)_{2L}\cr
     (N,D)_L+(N,N)_L&=(N,N)_{2L}\cr
     (N,N)_L+(D,D)_L&=P_{2L}\,,
     }
     \eql{relns}
  $$
where $P$ stands for periodic (\ie the generalised torus).

Hence for the mixed effective action,
  $$\eqalign{
      W^M_{I\times S^1}&=-{L\over12r}-\sum_{m=1}^\infty\bigg({1\over m(e^{4mL/r}-1)}-{1\over m(e^{2mL/r}-1)}\bigg)+{1\over2}\log2\cr
      &=-{L\over12r}+{1\over2}\sum_{m=1}^\infty {1\over m}\cosech 2mL/r+{1\over2}\log2\,,
  }
      \eql{wmixed}
  $$
a quantity I will return to. It is a function only of the shape, $L/r$, of the cylinder as the D and N conformal anomalies have cancelled.

Incidentally, from the mode relations, (\peq{relns}), it follows that the difference of $(D,D)$ and $(N,N)$ satisfies, quite generally,
  $$
  \big((N,N)-(D,D)\big)_L=\big((N,N)-(D,D)\big)_{2L}\,,
  $$
\ie is independent of the length of the cylinder, $\gC$, a fact reflected in the explicit quantities above. To determine what the value is, $L$ can be set to zero reducing $\gC$ to two copies of the cross--section manifold, $\man$, \ie the boundary of the cylinder.

Being a quantity of some significance and utility, (\peq{storus}) has been cast into various forms mostly associated with the modular properties of the torus. Possibly the quickest way to obtain these is to note the statistical mechanical equality (I set $r=1$ now),
  $$
  \sum_{m=1}^\infty{1\over m(e^{m\be}-1)}=-\log Q_0\,,
  $$
where,
  $$
  Q_0\equiv\prod_{n=1}^\infty(1-q^{2n})\,,\quad q=e^{-\be/2}=e^{-L}\,,
  $$
is the Euler product, used by Jacobi. A factor of $q^{1/12}$ takes us to the Dedekind $\eta$ function,
  $$
    \log\eta=\log Q_0-{\be\over24}\,.
  $$
Therefore, for the torus,
$$
  W_{S^1\times S^1}=2\log\eta(q)+\log2L\,.
  \eql{wtorus1}
  $$

I now return to the mixed effective action, (\peq{wmixed}), and rewrite this as a $q$--product,
  $$
     W^M_{I\times S^1}=-{L\over12}-\log Q_3(q^2)+{1\over2}\log2\,,
  $$
where $q$ is as above ($q=e^{-L}$) and,
  $$
    Q_3(q)\equiv\prod_{n=1}^\infty(1-q^{2n-1})\,.
  $$

Use of the standard identities,
  $$
   Q_3(q^2)=Q_2(q)\,Q_3(q)={1\over Q_1(q)}\,,
  $$
gives,
$$
     W^M_{I\times S^1}=-{L\over12}+\log Q_1(q)+{1\over2}\log2\,,
     \eql{wmixed5}
  $$
where
  $$
  Q_1(q)\equiv\prod_{n=1}^\infty(1+q^{2n})\,.
  $$
Reverting to a summation,
  $$
     W^M_{I\times S^1}=-{L\over12}-\sum_{m=1}^\infty{(-1)^m\over m}{1\over e^{2Lm}-1)}
     +{1\over2}\log2\,,
     \eql{wmixed2}
  $$
which corresponds to a scalar field quantised as a fermion, \ie $\oZ_2$ twisted around a thermal circle.

The $Q$ products can be written in terms of elliptic function quantities such as theta functions or the Weber class functions. For example, by definition,
  $$
         Q_1(q)={1\over\sqrt2}q^{-1/12}\gf_2(q)\,,
  $$
where $\gf_2$ is a Weber function, usually written $\gf_2(\tau)$ with $q=e^{\pi i\tau}$, the elliptic nome.

The factors in (\peq{wmixed5}) neatly combine  to give,
   $$
   W^M_{I\times S^1}=\log\gf_2(q)\,,
   \eql{wmixed3}
   $$
for the mixed, scalar effective action on the cylinder. ($\gf_2(q)$ is a monotonically decreasing function of $L$.)

There are many ways of deriving this result, all equivalent of course. For example, starting from (\peq{wtorus1}), and using (from (\peq{relns})),
  $$
  (N,D)_L={1\over2}\big(P_{4L}-P_{2L}\big)\,,
  $$
 the relation,
  $$
      \gf_2(q)=\sqrt2{\eta(q^2)\over\eta(q)}\,,
  $$
leads to (\peq{wmixed3}).

Under modular inversion, $q$ goes to $q'$, (the complementary nome). The Weber functions are related by,
  $$
    \gf_2(q')=\gf_1(q)\,,
  $$
and also by the doubling,
  $$
         \gf_1(q^2)\,\gf_2(q)=\sqrt2\,,
  $$
where $\gf_1$ is defined by,
  $$
   Q_3(q)=q^{1/24}\gf_1(q)\,.
  $$

Weber, [\pref{Weber2}], gives many special values for $\gf_1$ which can be used to find those for $\gf_2$. A few sample cases give the exact expressions,
$$\eqalign{
   W^M_{I\times S^1}&={1\over4}\log2\,,\quad L=\pi/\sqrt2\cr
   &={1\over6}\log2 +{1\over3}\log\big(\sqrt2(1+\sqrt5)\big)\,,\quad L=\pi/\sqrt{15}\cr
   &={3\over8}\log2\,,\quad L=\pi/2\,.
   }
   \eql{wmixed4}
   $$
\section{\bf 3. Low temperatures and the calculation of Pi}
When $L$ becomes large, the effective action is dominated by the vacuum energy, the first term in (\peq{wmixed2}). Equated to exact special values, this yields approximations for $e^\pi$ in terms of algebraic numbers.

The simplest course is to drop the summation in (\peq{wmixed2}). Exponentiation then gives,
     $$
         q^{1\over12}\approx{1\over\sqrt2}\gf_2(q)={1\over\gf_1(q^2)}\,,
     $$
and, redefining $q$ as $\sqrt{\ol q}$, produces the equation, ($\ol q\to0$),
    $$
    {\ol q}=\gf_1^{\,-24}(\ol q)\,,\quad{\rm q.p}.
    \eql{weber}
    $$
A particular choice, using an entry in Weber's list, [\pref{Weber2}], for $\ol q=e^{-\pi\sqrt{408}}\approx2.76025\times10^{-28}$ gives $e^\pi$ to 24 places from an algebraic quantity. The summation in (\peq{wmixed2}) can now be taken into account by an iterative procedure. The next approximation is easily shown to be,
   $$
    \ol q={1\over\gf^{\,24}_1(\ol q)}-{24\over \gf^{\,48}_1(\ol q)}\,,
    \eql{revers}
   $$
which doubles the precision. This process can be continued and is one example of a well--known approximation  scheme for $\pi$ using modular forms, [\pref{BandB}] \S 5.6. \footnote { Equation (\peq{revers}) corresponds to the first two terms of an expansion in $\gf_1^{-24}$ obtained by series reversion. }
\section{\bf 4. Fermions. Modes on a cylinder}
I now turn to free fermions on a cylinder of any dimension without specifying too closely the nature of the cross--section, $\man$, and proceed from the simplest boundary condition to the most complicated one. The ultimate aim is the evaluation of the functional determinant, and other spectral quantities on the way.

At any boundary, $\gB$, suitable, local conditions are provided by the `mixed' sort. These ensure self--adjointness of the Dirac operator, $D$, (and its square) and also that there is no flux across the boundary.  They are implemented by projection operators on the field. For example, mixed Dirichlet and Neumann (Robin) are,
  $$\eqalign{
  P_-\psi\,\big|_{{}_\gB}&=0\cr
 \big(n^\mu\nabla_\mu-{1\over2}\kappa\big)P_+\psi\,\big|_{{}_\gB}&=0\,,
 }
 \eql{mcond2}
  $$
where $\kappa$ is the extrinsic curvature of the boundary, and $n_\mu$ the inwards normal. $\nabla_\mu$ is the spinor covariant derivative and the projection operators are,
  $$
    P_{\pm}={1\over2}(1\pm i\ga\ga^\mu n_\mu)\,,
  $$
where $\ga$ is the generalisation of $\ga_5$.  These are the simplest, local conformally covariant conditions and, making the chiral split of the field,
  $$
      \psi=\bigg({\psi_+\atop \psi_-}\bigg)\,,
  $$
$P_-\psi=0$ gives, at the two ends of the cylinder,
  $$\eqalign{
  i\psi_-&=\psi_+\,,\quad z=0\cr
  i\psi_+&=\psi_-\,,\quad z=L\,,
       }
       \eql{mcond}
  $$
 where $z$ is the coordinate along the cylinder.

I describe the mode structure in more detail than usual as it is not quite standard \footnote{ This analysis is an extended version of that in [\pref{Apps}]. Another discussion is given in [\pref{KLP}] where several other references can be found.}. I have in mind the cylinder whose cross section is a $d$--sphere, but not exclusively. Formally any spin manifold could be used, so long as it lacks a zero mode and a spectral asymmetry. The spinor dimensions then are,
$$\eqalign{
    &2^{(d+1)/2}\,,\quad d\,\,{\rm odd}\cr
  &2^{(d+2)/2}\,,\quad d\,\,{\rm even}\,,
}
$$
and there is no boundary zero mode.\footnote{ One could refer to the boundary modes as Kaluza--Klein modes.}

The split Dirac equation reads,
  $$\eqalign{
 (\pa_z+{\widetilde D})\psi_-=-i\mu\psi_+\cr
 (\pa_z-{\widetilde D})\psi_+=-i\mu\psi_-\,,\cr
  }
  $$
 where $\widetilde D$ is the Dirac operator on S$^d$ for which the solutions are standard (I omit mode labels),
   $$
      \widetilde D\chi^{(\pm)}=\pm\la\chi^{(\pm)}\,.
   $$
On the $d$--sphere, $\la=\la_l=(l+d/2),\,l=0,1,\ldots$. It is not necessary to specify any degeneracies. These will be taken as understood in any boundary mode summations.

Separating variables, the eigenmodes of $D$ are, (with an implied spinor direct product),
  $$\eqalign{
  \bigg({\pm i|u|\atop u}\bigg)\,\chi^{(-)}e^{ikz}\,,\quad
  \bigg({\pm i|u|\atop u^*}\bigg)\,\chi^{(-)}e^{-ikz}\cr
  \bigg({\pm i|u|\atop -u^*}\bigg)\,\chi^{(+)}e^{ikz}\,,\quad
  \bigg({\pm i|u|\atop -u}\bigg)\,\chi^{(+)}e^{-ikz}\,,\cr
  }
  \eql{modes}
  $$
where $u=\la+ik$. To solve the equations, $\mu^2$ must equal $|u|^2$, giving the eigenvalues of $D$.\footnote{ The $\pm$ signs in the modes refers to the sign of the Dirac eigenvalue, $\mu$, on the cylinder.} Hence for $|u|$ one could just as well put $|\mu|$.

To accommodate (\peq{mcond}) it is necessary to take linear combinations of the modes, (\peq{modes}), \ie to mix left and right movers so giving standing waves.

One cannot mix $\chi^{(+)}$ with $\chi^{(-)}$ and the requisite combinations are,
  $$\eqalign{
  \psi^{(-)}_\pm&=(u^*\mp|u|)\bigg({\pm i|u|\atop u}\bigg)\,\chi^{(-)}e^{ikz}-
  (u\mp|u|)\bigg({\pm i|u|\atop u^*}\bigg)\,\chi^{(-)}e^{-ikz}\cr
  \psi^{(+)}_\pm&=(u\pm|u|)\bigg({\pm i|u|\atop -u^*}\bigg)\,\chi^{(+)}e^{ikz}-
  (u^*\pm|u|)\bigg({\pm i|u|\atop -u}\bigg)\,\chi^{(+)}e^{-ikz}\,,\cr
  }
  $$
which are better written in terms of standing waves, giving an expansion in the eigenstates of the projections, $P_{\pm}$, (see below), \footnote{ I generally drop any irrelevant overall constants.}
  $$\eqalign{
  \psi^{(-)}_\pm&=\pm k|\mu|\left({1\atop -i}\right)\chi^{(-)}\cos kz+|\mu|\big(|\mu|\pm\la\big)
  \left({1\atop i}\right)\chi^{(-)}\sin kz\cr
  \psi^{(+)}_\pm&=\mp k|\mu|\left({1\atop -i}\right)\chi^{(+)}\cos kz-|\mu|\big(|\mu|\pm\la\big)
  \left({1\atop i}\right)\chi^{(+)}\sin kz\,.\cr
  }
  \eql{swaves}
  $$

Conditions (\peq{mcond}) are satisfied at $z=0$. Applied at $z=L$ they  mean that the cosine term must vanish, thereby fixing $kL/\pi$ to be a half--integer, $\oZ_{-+}+1/2$, and determining the Dirac eigenvalues, $\mu$. The modes are anti--periodic with period $2L$.

Differentiating (\peq{swaves}) with respect to $z$ and setting $z$ to zero and to $L$ gives,
  $$\eqalign{
  i\psi'_+&=\psi'_-\,,\quad z=0\cr
  i\psi'_+&=\psi'_-\,,\quad z=L\,,
       }
       \eql{mcond3}
  $$
\ie the Neumann part of the conditions (\peq{mcond2}) is automatic, \cf\ [\pref{Luckock}], and this is typical.

It is helpful to give the representations of the relevant $\ga$--matrices. They can be regarded as two--dimensional,
  $$
  \ga^z=\left(\matrix{0&1\cr1&0}\right)\,,\quad  \ga=\left(\matrix{1&0\cr0&-1}\right)\,.
  $$

 Constructing the projections, $P_\pm$, at the boundary, $z=0$ and $z=L$, it is noticed that these are related by $P_\pm\big|_{z=0}=P_\mp\big|_{z=L}$ so all one needs to give is, say,
 $$
       P_{-}\big|_{z=L}={1\over2}\left(\matrix{1&i\cr-i&1}\right)\,,
 $$
with a null eigenvector, $\left({1\atop i}\right)$, and a unit one,  $\left({1\atop -i}\right)$. The boundary condition imposed above says that there should be no component proportional to the latter vector.

\section{\bf 5. Other boundary conditions}
For completeness I look at the condition obtained by interchanging $P_-$ and $P_+$
The easiest way of finding the modes is to use the fact that if $\psi$ satisfies the Dirac equation, then so does $\pa_z\psi$. This gives the combinations,
%\begin{ignore}
$$\eqalign{
  \psi^{(-)}_\pm&=\mp k|\mu|\left({1\atop -i}\right)\chi^{(-)}\sin kz+|\mu|\big(|\mu|\pm\la\big)
  \left({1\atop i}\right)\chi^{(-)}\cos kz\cr
  \psi^{(+)}_\pm&=\pm k|\mu|\left({1\atop -i}\right)\chi^{(+)}\sin kz-|\mu|\big(|\mu|\pm\la\big)
  \left({1\atop i}\right)\chi^{(+)}\cos kz\,.\cr
  }
  \eql{swaves2}
  $$
  %\end{ignore}
The condition $P_+\psi=0$ is satisfied at $z=0$ because   $\left({1\atop i}\right)$ is a null vector. It becomes a non--null vector of $P_+$ at $z=L$ and rejecting it
gives $kL/\pi=\oZ_++1/2$, as before.

These calculations have so far produced the Neveu-Schwarz eigenvalues that arise in string theory from anti--periodicity of the spinor around the `time' circle.

The periodic, Ramond eigenvalues can be obtained in the present approach by applying {\it different} conditions at the end points,\footnote{ Referred to as `separated' in [\pref{KLP}].}, say $P_+\psi=0$ at $z=0$ and $P_-\psi=0$ at $z=L$. (Observe that in this case the actual projection operators are the same.) The modes (\peq{swaves2}) satisfy the first condition, as just used. To satisfy the second one, the non--null vector, $\left({1\atop-i}\right)$, has to be removed so giving $kL/\pi=\oZ_{-+}$, \ie periodicity with period $2L$. The same conclusion is reached from the modes (\peq{swaves}) using the conjugate conditions, $P_-\psi=0$ at $z=0$ and $P_+\psi=0$ at $z=L$.

I note that the R and NS eigenvalues are here not obtained by requiring periodicity or anti--periodicity on a circle, but emerge from the finite cylinder. This has been encountered earlier, [\pref{BandS}], but not, seemingly, pursued.
\section{\bf 6. More general boundary conditions}
 The mode analysis is now extended to the more general mixed boundary conditions which are obtained from the projection, [\pref{Luckock}],
  $$
        P_\pm(\om)={1\over2}\big(1\pm n_\mu\ga^\mu\,e^{i\om\ga}\big)
        ={1\over2}\big(1\pm n_\mu\ga^\mu(\cos\om+i\ga\sin\om)\big)\,,
  $$
so that $P_\pm(\pi/2)$ are the $P_\pm$ used previously. The $\pm$ sign is really unnecessary as the sign can be changed by making $\om\to\pi+\om$. In the present geometry it is therefore sufficient just to look at,
  $$
  P_+(\om)\equiv{1\over2}\left(\matrix{1&e^{-i\om}\cr e^{i\om}&1}\right)\,,
  $$
with null ($\al$), and non--null ($\be$) eigenvectors, (normalised to 2), \footnote { I will not worry too much about overall normalisations.}
  $$
     \al=\left({1\atop-e^{i\om}}\right)\,,\quad \be=\left({1\atop e^{i\om}}\right)\,.
  $$

The construction of the relevant modes from (\peq{modes}) is straightforward. My  discussion is more particular and explicit than that in [\pref{BGKS}], for example.

To save energy, only the $\chi^{(-)}$ modes in (\peq{modes}) are considered and, further, since $\chi^{(-)}$ is a common factor, it will not be displayed. The generic mode thus takes the form,
  $$
  \Psi_{\pm}= A\,\bigg({\pm i|u|\atop u}\bigg)\,e^{ikz}+
  B\,\bigg({\pm i|u|\atop u^*}\bigg)\,e^{-ikz}\,,
  \eql{psi}
$$
which is to be written in terms of the projection eigenvectors,\footnote{ From now on, to save symbolism, the label indicating positive or negative Dirac eigenvalues, $\mu$, wll be taken as understood.}
  $$
  \Psi=X\al+Y\be\,.
  $$
The $X$ and $Y$ can be extracted from (\peq{psi}) by the ortho`normal' properies of the $\al$, $\be$,
  $$\eqalign{
     2X&=A\big(\pm i|u|-ue^{-i\om}\big)e^{ikz}+B\big(\pm i|u|-u^*e^{-i\om}\big)e^{-ikz}\cr
     2Y&=A\big(\pm i|u|+ue^{-i\om}\big)e^{ikz}+B\big(\pm i|u|+u^*e^{-i\om}\big)e^{-ikz}\,.\cr
     }
  $$

A boundary condition at $z=0$ can now be chosen by setting either $X$ or $Y$ to zero, say $Y=0$. This yields the ratio $A/B$ which can be substituted back into $X$ and $Y$ to give,
  $$\eqalign{
     Y&\propto \big(\pm|\mu|\sin\om-\la\big)\sin kz\cr
 X  &\propto\big(\pm i|\mu|\cos \om\sin kz+ k \cos kz\big)\,.\cr
  }
  \eql{modes4}
  $$
  This allows the effect of boundary conditions  at $z=L$ to be determined.

If one sets, again, $Y=0$, at $z=L$ the wave number $k$ takes Ramond values, $kL/\pi=\oZ_+$ quite simply,\footnote{ There are two sets of these values corresponding to the $\pm$ sign.} and no further consideration is required. This choice corresponds to applying {\it different} boundary conditions at the two ends. That is, $\om$ at one end and $\pi+\om$ at the other. (The sign change is actually cancelled by the change in sign of the normal vector.)

On the other hand, setting $X$ to 0 at $z=L$, yields a real $k^2$ (as demanded of a self--adjoint system) only
if $\om=\pi/2+ia$ with real $a$ \footnote{ These are `bag' boundary conditions investigated in some depth by [\pref{BGKS}] and also [\pref{EandK}], [\pref{EGK}].} , and then gives the {\it intrinsic} relation,
  $$
      \tan kL=\mp g{k\over\sqrt{\la^2+k^2}}\,,
      \eql{intrinsic}
  $$
where the parameter $g$ is defined by $g\equiv\cosech a$.

As a check, in the case $a=0$, one recovers (two sets of) the NS half--integer eigenvalues, $k=\oZ_{-+}+1/2$ while if $a=\infty$ there result (two sets of) R eigenvalues. Between these two limits one has what might be called, `interpolating' eigenvalues.
Equivalent statements hold for the $\chi^{(+)}$ modes.

From (\peq{intrinsic}), it can be seen that there is an asymmetry between $\mu$--positive and $\mu$--negative cylinder modes.\footnote{ This has been noticed earlier and analysed by Beneventano {\it et al} [\pref{BSW}] who set APS spectral conditions at one end and local bag at the other.  My discussion therefore differs somewhat and is rather more detailed.} Furthermore, for $0<g<\la L$, and $\mu>0$ there exists an imaginary $k$--root, $k=i\ga$. The same holds for $0>g>-\la L$ and $\mu<0$.
The total eigenvalue, $\mu$, on the cylinder is still real. For a given value of $g$, there will be a an infinite number of these `imaginary' $k$--modes (one per qualifying $\la$) with $\mu^2=\la^2-\ga^2>0$.

If $g=\pm\la L$, there is a true $k=0$ root, in which case the modes are linear functions of $z$. That is, $Y\sim z$ and $X\sim L-z$ with normalisation factors of $\sqrt3/L^{3/2}$. Of course, these modes must be retained as part of the mode structure on the cylinder, having $\mu=\pm\la$.

In the limit $g\to \infty$, the imaginary root tends to zero ($k=0$) and, in the Ramond limit of $g\to0$, it corresponds to a zero mode of the Dirac operator on the cylinder, \ie $\mu=0$. This must be omitted in spectral quantities.

Equation (\peq{intrinsic}) determines the wave number, $k$, as a function of the boundary eigenvalues, $\la$. Since, in the present work, I am not concerned with the spectral asymmetry, it suffices to remove the radical by squaring, to give the Mittag--Leffler/Rayleigh /Euler form,
    $$
 1-  \big(\la^2\sinh^2a+\xi^2\cosh^2a\big){\sin^2 \xi L\over \xi^2}
   =\big(1-\la^2L^2\sinh^2a\big)\prod_{k}\big(1-{\xi^2\over k^2}\big)\,,
   \eql{ML}
   $$
where the product is over all non--zero roots of (\peq{intrinsic}), for a fixed $\la$ {\footnote{ There are still the two sets of roots straddling the points $\oZ_{-+}+1/2$.}.

A few consequences of these eigenvalue systems are explored in section 10.
\section{\bf 7. The most general boundary condition}
The boundary conditions employed in the preceding section are not the most general as the parameter $a$ can take any value at an end point.

In general, therefore, there are two unrelated projection operators,
$$
  P_+(a_j)\equiv{1\over2}\left(\matrix{1&-ie^{-a_j}\cr ie^{a_j}&1}\right)\,,
  \quad j=0,1
  $$
which can be applied at the cylinder ends. The eigenmode is still given by (\peq{psi}) and can be expanded in either of the two sets of projection eigenvectors,
$$
  \Psi=X_j\al_j+Y_j\be_j\,,
  \eql{gmode}
  $$
with coefficients,
$$\eqalign{
    2 X_j&=A\big(\pm |u|+ue^{-a_j}\big)e^{ikz}+B\big(\pm |u|+u^*e^{-a_j}\big)e^{-ikz}\cr
     2Y_j&=A\big(\pm |u|-ue^{-a_j}\big)e^{ikz}+B\big(\pm |u|-u^*e^{-a_j}\big)e^{-ikz}\,.\cr
     }
  $$

Without loss of generality, the boundary condition is set to be $Y_0$=0 at $z=0$ and so,
   $$
         B=-A{\pm |u|-ue^{-a_0}\over \pm |u|-u^*e^{-a_0}}\,.
   $$
Then, after slight algebra, the modes become, {\it in this case},
$$\eqalign{
X_0&\propto\pm|u|\sinh a_0\sin kz+ k\cos kz\cr
Y_0&\propto(\pm|u|\cosh a_0-\la\big)\sin kz \cr
X_1&\propto\big(\pm|u|\sinh a-\la \sinh b\big)\sin kz+ k\cosh b\cos kz\cr
Y_1&\propto\big(\pm|u|\cosh a -\la \cosh b \big)\sin kz+ k\sinh b\cos kz\,.
}
\eql{modes3}
$$

\begin{ignore}
$$\eqalign{
     X_j&\propto \big(\pm |u|-u^*e^{-a_0}\big) \big(\pm |u|+ue^{-a_j}\big)e^{ikz}\cr
     &\hspace{***************}-\big({\pm |u|-ue^{-a_0}}\big)\big(\pm |u|+u^*e^{-a_j}\big)e^{-ikz}\cr
     Y_j&\propto \big(\pm |u|-u^*e^{-a_0}\big)\big(\pm |u|-ue^{-a_j}\big)e^{ikz}\cr
     &\hspace{***************}-\big(\pm |u|-ue^{-a_0}\big)\big(\pm |u|-u^*e^{-a_j}\big)e^{-ikz}\,.\cr
     }
  $$
multiplying out the brackets, and dropping overall factors,
$$\eqalign{
     X_j&\propto \big(|u|-|u|e^{-a_0-a_j}\mp u^*e^{-a_0}\pm ue^{-a_j}\big)e^{ikz}\cr
     &\hspace{**}-\big(|u|-|u|e^{-a_j-a_0}\mp ue^{-a_0}\pm u^* e^{-a_j}\big)e^{-ikz}\cr
     Y_j&\propto \big(|u|+|u|e^{-a_0-a_j}\mp u^*e^{-a_0}\mp ue^{-a_j}\big)e^{ikz}\cr
     &\hspace{**}-\big(|u|+|u|e^{-a_j-a_0}\mp ue^{-a_0}\mp u^* e^{-a_j}\big)e^{-ikz}\cr
     }
  $$
therefore
$$\eqalign{
     X_j&\propto 2i|u|\big(1-e^{-a_0-a_j}\big)\sin kz +\big(\mp u^*e^{-a_0}\pm ue^{-a_j}-\big(\mp ue^{-a_0}\pm u^* e^{-a_j}\big)\big)(\cos kz)\cr
     &\hspace{*******************}+(\big(\mp u^*e^{-a_0}\pm ue^{-a_j})+\big(\mp ue^{-a_0}\pm u^* e^{-a_j}\big)i\sin kz\cr
     &\propto 2i|u|\big(1-e^{-a_0-a_j}\big)\sin kz -\big(\pm u^*e^{-a_0}\mp  ue^{-a_j}\mp  ue^{-a_0}\pm u^* e^{-a_j}\big)\cos kz\cr
     &\hspace{*******************}-\big(\pm u^*e^{-a_0}\mp  ue^{-a_j}\pm  ue^{-a_0}\mp u^* e^{-a_j}\big)i\sin kz\cr
      &\propto |u|\big(1-e^{-a_0-a_j}\big)\sin kz \pm k\big(e^{-a_0}+ e^{-a_j}\big)\cos kz\cr
     &\hspace{*******************}\mp \la\big(e^{-a_0}- e^{-a_j}\big)\sin kz\cr
     Y_j&\propto \big(|u|+|u|e^{-a_0-a_j}\mp u^*e^{-a_0}\mp ue^{-a_j}\big)e^{ikz}\cr
     &\hspace{**}-\big(|u|+|u|e^{-a_j-a_0}\mp ue^{-a_0}\mp u^* e^{-a_j}\big)e^{-ikz}\cr
     &\propto 2i|u|(1+e^{-a_0-a_j})\sin kz+(\mp u^*e^{-a_0}\mp ue^{-a_j}\pm ue^{-a_0}\pm u^*e^{-a_j})\cos kz\cr
     &\hspace{*************}+(\mp u^*e^{-a_0}\mp ue^{-a_j}\mp ue^{-a_0}\mp u^*e^{-a_j})i\sin kz\cr
     &\propto|u|(1+e^{-a_0-a_j})\sin kz\pm k(e^{-a_0}-e^{-a_j})\cos kz\mp\la ( e^{-a_0}+e^{-a_j})\sin kz\,.
     }
  $$
 \end{ignore}
New parameters have been defined by $a=(a_0+a_1)/2$ and $b=(a_1-a_0)/2$ so that,
  $$
  {e^{-a_0}\pm e^{-a_1}\over  1-e^{-a_0-a_1   }}=
  {\cosh b\atop\sinh b}\bigg\}\cosech a\quad{\rm and}\quad {e^{-a_0}\pm e^{-a_1}\over  1+e^{-a_0-a_1   }}=
  {\cosh b\atop\sinh b}\bigg\}\sech a\,.
  $$

  The eigenvalues are determined by selecting a boundary condition at $z=L$.
  It is seen that the forms of $X_0$ and $Y_0$ agree with those in the simpler situation of the previous section, (\peq{modes4}), with $a$ there replaced by $a_0$ here so that if $X_0(L)$ is set to zero, the previous results will be replicated, completely independent of the value of $a_1$.

  A more complicated outcome follows on the choice $X_1(L)=0$, which produces the intrinsic relation,
$$\eqalign{
  \big(\pm\sqrt{\la^2+k^2}\,\sinh a-\la\sinh b\big){\sin kL\over k} + \cosh b\cos kL=0\,,\cr
  }
  \eql{intrinsic4}
  $$
and I concentrate on this for the rest of this paper. The condition $Y_1(L)=0$ will give analogous results.

The  expression obtained from the $k\to 0$ limit\footnote{ One can also take the derivative.} of (\peq{intrinsic4}) yields the condition for the existence of imaginary $k$--modes,
  $$\eqalign{
    \la L>{\cosh b\over \sinh b\mp\sinh a}=\bigg\{{{1\over2}\big(t_1-t^{-1}_0\big)\atop
    {1\over2}\big(t^{-1}_1-t_0\big)}\,,
  }
  \eql{imagcond}
  $$
where the, here, more handy parameters, are, $t_0\equiv\tanh a_0/2$ and $t_1\equiv \tanh a_1/2\,,$ which run, monotonically, between $-1$ and $+1$.

The number of imaginary $k$--roots  depends on the distribution of the parameters. Divide the $t$--parameter domain (a square) into quadrants,
  $$\eqalign{
  \caD_1=&\{0<t_0<1\,,-1<t_1<0\}\,,\quad \caD_2= \{-1<t_0<0\,,-1<t_1<0\}\,,\quad \cr
  \caD_3= &\{0<t_0<1\,,0<t_1<1\}\,,\quad  \caD_4=\{-1<t_0<0\,,0<t_1<1\}\,.
  }
  $$
In $\caD_4$, at qualifying values of $\la$, there will be an imaginary $k$--root for both positive and negative Dirac eigenvalues, $\mu$, but in $\caD_3$ and $\caD_2$ only for positive and negative $\mu$, respectively. In $\caD_1$ there are no imaginary roots.

As $t_0$ and $t_1$ range over the parameter space, imaginary roots can disappear
and re--appear as real ones (and {\it vice versa})  going via zero roots. This behaviour can be followed graphically in much the same way as elaborated in Strauss, [\pref{Strauss}], for scalar Robin conditions on an interval.
\section{\bf 8. Imaginary modes for  large L}
The existence of imaginary roots raises the subsidiary question of the behaviour as the length of the cylinder, $L$, tends to infinity since these modes involve real exponentials. \footnote{ This is sometimes called the adiabatic limit.} I investigate them a little more, with a curious result.

 The equation for an imaginary root, $k=i\ga\,$($\ga>0$) is,
$$\eqalign{
  \big(\pm\sinh a\sqrt{\la^2-\ga^2}-\la\sinh b\big){\tanh \ga L\over \ga} + \cosh b=0\,,\cr
  }
  \eql{imagr}
  $$
which, as $L\to\infty$ becomes,
$$\eqalign{
 \pm \sqrt{\la^2-\ga^2}\sinh a- \la\sinh b +\ga\cosh b=0\,.\cr
  }
  \eql{imagr1}
  $$
  Solving the quadratic produces two possible roots,
  $$
   \ga_+=-\la\tanh a_0\,\quad{\rm and}\quad\, \ga_-=\la\tanh a_1\,,
  $$
on the  positive and negative $\mu$ branches, respectively.

These values accord with the root distribution encountered earlier. Because $\ga$ is restricted to be positive, there are two roots if $a_0<0$ and $a_1>0$, no roots if $a_0>0$ and $a_1<0$, one (positive branch) root if $a_0<0$ and $a_1<0$ and one (negative branch) root if $a_0>0$ and $a_1>0$.

To see what the mode (\peq{gmode}) becomes in this limit, the $X_1$, $Y_1$ set is the most relevant and the exponential forms most appropriate.

\begin{ignore}
$X_0$ and $Y_0$ are,
  $$\eqalign{
X_0&\propto|u|\sinh a_0\sinh \ga_+ z+ \ga_+\cosh \ga_+ z\cr
&\propto {1\over\cosh a_1}\sinh a_0\sinh \ga_+ z-\tanh a_1\cosh \ga_+ z \cr
Y_0&\propto(|u|\cosh a_0-\ga_+\big)\sinh \ga_+ z \cr
&\propto\big({1\over\cosh a_1}\cosh a_0-1\big)\sinh\ga_+z\cr
&
}
\eql{modes3}
$$
\end{ignore}
The continuation $k\to i\ga$ gives,
$$
     |u|\to\sqrt{\la^2-\ga^2}\,,\quad u\to\la-\ga\,,\quad u^*\to\la+\ga\,.
$$
Then, for the $X$--modes,
$$\eqalign{
     &X_j\propto\big(|u|(1-e^{-a_0-a_j})\mp \la(e^{-a_0}-e^{-a_j})\mp\ga(e^{-a_0}+ e^{-a_j})\big)e^{-\ga z}\cr
     &\hspace{**}-\big(|u|(1-e^{-a_j-a_0})\mp \la (e^{-a_0}-e^{-a_j})\pm \ga(e^{-a_o}+ e^{-a_j})\big)e^{\ga z}\,,\cr
     }
     $$
 so that,
     $$\eqalign{
    X_1 &\propto \big(|u|\sinh a\mp\la\sinh b\mp\ga\cosh b\big)e^{-\ga z}
     -\big(|u|\sinh a\mp\la\sinh b\pm\ga\cosh b\big)e^{\ga z}\,.\cr
     }
     $$
For the $Y$--modes,
     $$\eqalign{
     Y_j&\propto \big(|u|(1+e^{-a_0-a_j})\mp \la(e^{-a_0}+e^{a_j})\mp\ga(e^{-a_0}-e^{-a_j})\big)e^{-\ga z}\cr
     &\hspace{**}-\big(|u|(1+e^{-a_j-a_0})\mp \la (e^{-a_0}+e^{-a_j})\pm \ga(e^{-a_0}- e^{-a_j})\big)e^{\ga z}\,,\cr
     }
     $$
and so,
     $$\eqalign{
     &Y_1\propto\big(|u|\cosh a\mp \la\cosh b\mp\ga\sinh b\big)e^{-\ga z}-\big(|u|\cosh a\mp \la\cosh b\pm \ga\sinh b\big)e^{\ga z}\,.\cr
     }
  $$
The roots on the positive and negative branches, for large $L$, \ie,
  $$
  \ga_+\approx-\la\tanh a_0\,\quad{\rm and}\quad \ga_-\approx\la\tanh a_1\,,
  $$
give for the absolute values of the corresponding Dirac eigenvalues,
$$
  |u|_+\approx{\la\over \cosh a_0}\,\quad {\rm and}\quad|u|_-\approx{\la\over\cosh a_1}\,.
$$
Substitution of these into the modes above gives, on the positive branch, after cancellations using $b+a_0=a$ and $b-a_1=-a$,
$$\eqalign{
Y_1^+&\propto e^{-\ga_+z}\cr
X^{+}_1&\propto e^{-\ga_+z}\,,\cr
}
$$
and on the negative branch,
$$\eqalign{
Y_1^-&\propto\cosh b\cosh a_1\, e^{-\ga_-z}-\cosh a \,e^{\ga_-z}\,\cr
     X^{-}_1&\propto e^{-\ga_-z}\,.
}
$$

There is an apparent serious asymmetry between positive and negative Dirac eigenvalues in that $Y_1^-$ diverges as $z$ tends to infinity. For large, but finite, $L$, this would seem to be an allowable mode.

I have not been able to resolve this difficulty (if it is one) with the $L\to\infty$ limit.

\section{\bf 9. The squared formulation}

As before, unless one is interested in the spectral asymmetry, it is sufficient to
eliminate the radical thus furnishing the squared form of (\peq{intrinsic4}),
\begin{ignore}
$$\eqalign{
  (\la^2+k^2){\sin^2 kL\over k^2} -\bigg(\cosh b\cos kL- \la\sinh b\,{\sin kL\over k}\bigg)^2\cosech^2 a
  }
  \eql{intrinsic3}
  $$
\end{ignore}
with the corresponding Euler relation,\footnote{ This is left in a cosmetically unsimplified
 form.}  $$\eqalign{
  (\la^2+\xi^2){\sin^2 \xi L\over \xi^2}& -\bigg(\cosh b\cos \xi L- \la\sinh b\,{\sin \xi L\over \xi}\bigg)^2\cosech^2 a\cr
  &=(\la L-l_+)(\la L-l_-)\prod_{k}\big(1-{\xi^2\over k^2}\big)\,,
  }
  \eql{ML2}
  $$
where $l_\pm$ are the roots of a quadratic,
  $$
           l_\pm={\cosh b\over\sinh b\pm\sinh a}\,,
  $$
  that appear also in (\peq{imagcond}).

The logdet for this system is evaluated in section 12.

\section{\bf 10. Spinor spectral matters. The determinant}
I am particularly interested in the logdet of the system with interpolating eigenvalues.  This is defined to be $-\ze'_\gC(0)$ in terms of the \zf\ of the (squared) Dirac operator (defined below) and involves a sum over all the eigenvalues. As an ansatz, I begin by working at a fixed value of the boundary eigenvalue, $\la$, which is treated as a spectator. The sum over $k$ then behaves like a one--dimensional problem.

In order to extract spectral quantities a parameter, $m$, is introduced by giving the Dirac field a mass, through adding $\kappa^2$ to $\la^2$, \ie $\la^2\to \la^2+\kappa^2$, which will be denoted $m^2$. (It is handy to have both $\ka$ \footnote { Not to be confused with the extrinsic curvature.}  and $m$ available.)

For the time being, the boundary manifold is forgotten  by defining a subsidiary, artificial one--dimensional \zf,
   $$
        \ze(s,m^2)=\sum_{k}{1\over (k^2+m^2)^s}\,,
        \eql{kzeta}
   $$
where the sum is over all  the non--zero roots of the intrinsic equation (\peq{intrinsic}), for a fixed $\la$. (The roots therefore depend on $\la$.)

For simplicity, discussion is initially restricted to the case of (\peq{intrinsic}) and,
a comparison with (\peq{ML}) shows that $\xi=im$ and that the resolvent, $R$,  is,
    $$\eqalign{
    R(m^2)&\equiv\ze(1,m^2)={d\over d m^2}\log\bigg(1+\big(m^2\cosh^2a-\la^2\sinh^2a\big){\sinh^2mL\over m^2}\bigg)\cr
    &\equiv{d\over d m^2}\log F(m^2)\,.
    }
    \eql{resolv}
$$

Although it is an artificial, intermediate stage, it is interesting to extract the summations of inverse powers of the roots of the intrinsic equation by expanding in $m^2$. For example, one obtains the Rayleigh--type sum--rules, [\pref{Rayleigh}],\footnote{ Otherwise known as trace identities.}
  $$\eqalign{
  \sum_k{1\over k^2}&=L^2{3 +3 g^2 -L^2 \lambda^2
  \over3 (g^2-L^2 \lambda^2)}\cr
  \sum_k{1\over k^4}&=L^4{45 +60 g^2 +15 g^4 +4 g^2 L^2\lambda^2+L^4 \lambda^4\over45 (g^2-L^2 \lambda^2)^2}\,,
  }
  $$
which can be checked numerically. The sum includes the negative mode, if it exists. As $g^2\to\la^2 L^2$  it dominates these sums. At $g^2=\la^2 L^2$ there is a zero mode (of this one--dimensional sector).

Expanding in powers of $\kappa^2$, which is the relevant parameter for the complete cylinder, gives, more interestingly,
  $$\eqalign{
  \sum_k{1\over k^2+\la^2}&=
  \tanh(L \lambda)\, {g^2 L \lambda+\tanh(L\lambda)\over g^2 \lambda^2}\cr
   \sum_k{1\over (k^2+\la^2)^2%%
   }&=-{1\over2g^4\la^4}\bigg(g^4\la^2L^2\sech^2\la L+g^2\la L(4\sech^2\la L-g^2)\tanh\la L\cr
   &-2\tanh^2\la L\big(2g^2+\tanh^2\la L\big)
   \bigg)\,,
   }
   \eql{sr}
  $$
  \vskip 3truept
\noindent and so on.  Again, all $k$--modes {\it must} be included. Numerical evaluation verifies these relations in which $k$ depends on $\la$. \footnote{ For the values $\la=3/2$, $g=2$, $L=1$, sixty terms in the second sum--rule give 0.1589115 while the right--hand--side (the exact value) is 0.1589118. When an imaginary root exists, say for $\la=5/2$, $g=2$, $L=1$, the numbers are (0.0113596+ 0.0312205)=0.0425801 and 0.0425803.} Note that these quantities are polynomials in the boundary condition parameter, $1/g$.

Depending on the dimension of the boundary, further summation  over $\la$ would give the values of the complete cylinder \zf, $\ze_\gC(s)$, (see below)  at positive integer $s$. For a $d$--dimensional cylinder, $s=d,d+1,\ldots$. Equation (\peq{sr}) could then be used for numerical calculation but, as shown below, the convergence can be improved and the range of validity increased by a straightforward manoeuvre.

To explain this, I generalise these sums to the  \zf\ of the  squared Dirac operator on the cylinder, at a fixed value of $\la$,
   $$
         \ze_\gC(s,\la)=\sum_k {1\over (k^2+\la^2)^s}\,.
         \eql{parze}
   $$
Further summation over $\la$ produces the \zf\ on the cylinder,
  $$
     \ze_\gC(s)\equiv\sum_\la \ze_\gC(s,\la),
     \eql{totze}
  $$
the sum being over all $\la$, distinct or not. This is the object of ultimate interest. Its construction is given later.

The argument $s$ has to be in a region for which the double summation converges, in principle. However, the sum--rules, (\peq{sr}), can be adapted to give the values of the total \zf\ at positive integers at all $d$ in the following, not unknown, way. They are regularised, {\it \`{a} la} Weierstrass, by, first, subtracting and adding the large $\la$ behaviour.  and then summing the latter over $\la$ in terms of the boundary \zf, defined by,
   $$
         \ze_\man(s)=\sum_\la {1\over\la^{2s}}\,,
   $$
whose continuation is assumed known. I illustrate this process in the simplest case of $s=n=1$,

Performing the regularisation (denoted by $*$), gives,
  $$\eqalign{
 {\sum_{k.\la}}^*{1\over k^2+\la^2}&=L\sum_\la{\tanh\la L-1\over \la}-
{1\over g^2} \sum_\la{\sech^2\la L\over \la^2}+L\,\ze_\man(1/2)+{1\over g^2}\ze_\man(1)\,,\cr
 }
 \eql{regsr}
  $$
which can easily be evaluated numerically, given the spectrum $\la$ and $\ze_\man$. It will be confirmed later.

Turning now to the determinant, I follow the direct method advocated in [\pref{KLP}]. This is based on a Cauchy residue theorem the rearrangement of which gives an integral form for the partial \zf, $\ze_\gC(s,\la)$,
as,
  $$
    \ze_\gC(s,\la)={\sin\pi s\over s}\int_0^\infty d\ka\,\ka^{-2s} {d\over d\ka}\log F(\ka^2+\la^2)\,,
  $$
where $F$  is given in (\peq{resolv}).

The expression for $F$ is arranged in a fashion suitable for integration in a way similar to a related computation in [\pref{KLP}]. Thus (I retain $m=(\la^2+\ka^2)^{1/2}$ for ease of writing),
   $$\eqalign{
       F(m^2)={e^{2mL}\cosh^2a \over4(\ka^2+\la^2)}\big(\ka^2+{\la^2\over
       \cosh^2a}\big)
       \bigg(\!1+2{\ka^2(1-\sinh^2a)+\la^2\over \ka^2\cosh^2a+\la^2}\,e^{-2mL}+e^{-4mL}\!\bigg)\,.\cr
       }
       \eql{eff}
   $$
This structure corresponds to removing the large $\la$ behaviour, as used earlier.

There are four distinct contributions to the integrated logarithmic derivative corresponding to the four factors in (\peq{eff}), barring constant multipliers. All the resulting integrals have usefully been given in the form required in [\pref{KLP}] and for convenience are copied here,
  $$\eqalign{
        {\sin\pi s\over\pi}\int_0^\infty d\ka\,\ka^{-2s}{d\over d\ka}\log e^{Lm}
        &={L\over\la^{2s-1}}{\Ga(s-1/2)\over2\sqrt\pi\Ga(s)}\cr
        {\sin\pi s\over\pi}\int_0^\infty d\ka\,\ka^{-2s}{d\over d\ka}\log(\ka^2+\la^2)
        &={1\over\la^{2s}}\,.
        }
  $$
These are the only integrals that can be, and have to be, done explicitly in order to effect the necessary continuation in $s$. Using them, one derives, upon summing over the boundary eigenvalues,
   $$\eqalign{
        \ze_\gC(s)&=2L\,\ze_\man(s-1/2){\Ga(s-1/2)\over2\sqrt\pi\Ga(s)}
        +(\cosh^{2s} a-1)\ze_\man(s)\cr
        &+{\sin\pi s\over\pi}\sum_\la \int_0^\infty \ka^{-2s}{d\over d\ka}
       \log \bigg(1+
       2{\ka^2(1-\sinh^2a)+\la^2\over \ka^2\cosh^2a+\la^2}\,e^{-2mL}+e^{-4mL}\bigg)\,.
        }
        \eql{cylzeta}
   $$

This is as far as one can go without explicit knowledge of the boundary spectral data. Useful information can, however, be gained at specific values of $s$. For example, for the logdet it follows that,
   $$\eqalign{
    \log\det D^2=&-2\log\cosh a\,\ze_\man(0)+2L\,\ze_\man(-1/2)\cr
    &+\sum_\la\log \bigg(1+
       2{\ka^2(1-\sinh^2a)+\la^2\over \ka^2\cosh^2a+\la^2}\,e^{-2mL}+e^{-4mL}\bigg)\,\bigg|^\infty_0\cr
    =&-2\log\cosh a\,\ze_\man(0)+2L\ze_\man(-1/2)+
    2\sum_\la\log(1+e^{-2\la L})\,.
    }
    \eql{logdet}
   $$
This should be multiplied by a factor of two to account for the $\chi^{(+)}$ boundary modes which, as is easily shown, contribute an equal amount.

For the NS condition of $a=0$, the dependence on the boundary conformal anomaly, $\ze_\man(0)$, disappears corresponding to the fact that these conditions contain equal amounts of Neumann and Dirichlet.  For the Ramond condition of $a=\infty$, the divergence due to the overall zero mode ($\mu=0$) must be removed.

If the boundary is closed and odd dimensional, $\ze_\man(0)$ vanishes, assuming no zero modes, and the dependence on the boundary conditions disappears in this case too. For conformal spinors on spheres, this is always the case and so the logdet is independent of the boundary condition on the cylinder,\mgn{Is this really correct?} which is, perhaps, a little unexpected. This situation changes for the most general boundary conditions, as detailed in section 12.

The last two terms in (\peq{logdet}) have the standard statistical mechanical form with $\sim \ze_\man(-1/2)$ the boundary vacuum energy.

I note that, unlike a similar calculation in [\pref{KLP}] (but for a different boundary condition), there is no dependence on the boundary logdet, $-\ze_\man'(0)$.

The form (\peq{logdet}) allows other values to be obtained, in particular those when $s$ is integral. For example the cylinder conformal anomaly, $\ze_\gC(0)$, always vanishes.

To make contact with the values obtained earlier from the regularised sum--rules, it is necessary, when $s$ is positive integral,  to analyse the last term more closely. As usual, the integration is split,
  $$
   \int_0^1 d\kappa \,\kappa^{-2s} \La({\kappa})+
   \int_1^\infty d\kappa\, \kappa^{-2s} \La({\kappa})\,,
  $$
where $\La(\kappa)$ is the remaining factor in the integrand. The second integral converges
and so, multiplied by $\sin\pi s$ vanishes on the integers. The first integral is investigated by expanding $\La(\kappa)$ in (odd) powers of  $\kappa$,
   $$
      \La(\kappa)=2\La_1\ka+2\La_2\ka^3+\ldots\,,
   $$
where the coefficients are functions of $\la$ (and $g$ and $L$) and using the continuation,
   $$
        \int_0^1{1\over \ka^{2s-2n+1}}=-{1\over 2(s-n)}\,,
   $$
so that,
  $$
  \int_0^1 d\ka \La(\ka)=-{\La_1\over s-1}-{\La_2\over s-2}-\,\ldots\,.
  $$

Evaluating the cylinder \zf, (\peq{cylzeta}) at $s=n\in\oZ_+$ then  gives,
  $$
      \ze_\gC(n)=L\,\ze_\man(n-1/2){\Ga(n-1/2)\over \sqrt \pi (n-1)!}+\bigg(\cosh ^{2n}a-1\bigg)\ze_\man(n)-
      (-1)^n\sum_\la \La_n(\la,g,L) \,.
  $$

When $n=1$,
$$
      \ze_\gC(1)={L}\,\ze_\man(1/2)+\sinh^2a\,\ze_\man(1)+\sum_\la \La_1(\la,g,L) \,.
  $$
$$
\La_1={2\over g^2\la^2}\,\sech^2(\lambda  L)
+{{2L}\over{\lambda }}(1-\tanh\la L)\,,
$$
which agrees with the result found earlier, (\peq{regsr}).

For $n=2$, I just record an expression for the expansion coefficient, $\La_2$,
 $$ \eqalign{
-\La_2=&16{{\left(e^{2 \lambda  L}+ e^{6 \lambda  L}\right)}\over
{g^4\lambda ^4 \left(e^{2 \lambda  L}+1\right)^4}}+
16{{\left( e^{6 \lambda  L} (\lambda  L+1)+2 e^{4 \lambda  L}+e^{2 \lambda  L}
(1- \lambda  L)\right)}\over{g^2\lambda ^4 \left(e^{2 \lambda  L}+1\right)^4}}\cr
&+2L{{ 1+  e^{6 \lambda  L} (2 \lambda  L+1)+ e^{2 \lambda  L} (2 \lambda  L+3)+ e^{4 \lambda  L} (4 \lambda  L+3)}\over{\lambda ^3 \left(e^{2 \lambda  L}+1\right)^4}}\,,\cr
}
$$
which enters into the \zf\ value,
$$
      \ze_\gC(2)={L\over2}\,\ze_\man(3/2)+\big(\cosh^4a-1\big)\ze_\man(2)-\sum_\la \La_2(\la,g,L) \,.
  $$

I note that similar algebraic factors occur in the more explicit parts of [\pref{EGK}] and [\pref{EandK}].

A very general way of computing \zf\ values for Sturm-Liouville problems using contour techniques  is contained in [\pref{FGKS}] where many other references can be found.

\section{\bf 11. Another computation of the logdet}

I now give an alternative, dangerous, but fast, derivation of the logdet based on
the classic way of analysing spectral quantities, particularly resolvents and infinite products, by examination of the limit $m\to \infty$.\footnote{ This technique has been used for a similar situation in [\pref{dowrobin}]. Canonical products also early appear in [\pref{Death}] for local spinor boundary conditions in spherical geometries.} This bypasses, in the present instance, the need for continuation at the expense of an {\it ad hoc} recipe, in similar vein to the Weierstrass regularisation employed earlier.

Firstly, for the logdet, from (\peq{ML}), one has the formal relation,
   $$
   \ze'(0,\la^2)=\ze'(0,m^2)+\log F(m^2)-\log F(\la^2)\,,
   $$
for all $m$. Therefore, in particular,
$$
    \ze'(0,\la^2)=\lim_{m\to\infty}\big(\ze'(0,m^2)+\log F(m^2)\big)-\log F(\la^2)\,,
    \eql{det}
$$
requiring knowledge of the first term on the right--hand--side. This can be obtained from the asymptotic form of $\ze(s,m^2)$ derived long ago in terms of (here, one--dimensional)  `heat--kernel' coefficients which can be found by comparing this asymptotic form with that of the {\it explicit} resolvent, (\peq{resolv}). An elementary calculation then reveals that the limit in (\peq{det}) is $2\log({1\over2}\cosh a)$ and so, for the logdet of this one--dimensional system, there results,\footnote { There should be another factor of two coming from the boundary $\chi^{(+)}$ modes.}
$$\eqalign{
    -\ze'(0,\la^2)&=\log F(\la^2)-2\log({1\over2}\cosh a)\cr
    &=-2\log\cosh a+2\la L +2\log\big(1+e^{-2\la L}\big)\,.
    }
    \eql{det2}
$$

The full logdet on the cylinder is obtained by summing the contributions for each $\la$. The final term in (\peq{det2}) converges and needs no further action. To deal with the other terms, I invoke a further ansatz by making the replacements,
    $$
         \sum_\la 1 \Rightarrow \ze_\man(0)\,,\quad \sum_\la \la\Rightarrow\ze_\man(-1/2)\,,
    $$
corresponding to the boundary conformal anomaly and vacuum energy.

The complete cylinder logdet then reads,
   $$
    \log\det D^2=-2\log\cosh a\,\,\ze_\man(0)
    +2L\,\ze_\man(-1/2) +2\sum_\la\log\big(1+e^{-2\la L}\big)\,.
   $$
in agreement with (\peq{logdet}) with little effort.\footnote{ More sophisticated treatments of this general method of calculation are available, \eg [\pref{Voros}].}

\section{\bf 12. logdet for general conditions}

For the most general bag conditions, the procedures remain the same, only  the function, $F(m^2)$, in the resolvent changes to, \cf (\peq{ML2}),
$$\eqalign{
 F(m^2)= (m^2-\la^2){\sinh^2 m L\over m^2}& +\bigg(\cosh b\cosh m L- \la\sinh b\,{\sinh m L\over m}\bigg)^2\cosech^2 a\,.\cr
  }
  \eql{eff2}
  $$

For simplicity I consider just the determinant, which, for speed, is computed by the method of the previous section using equation (\peq{det}).

The large $m$ behaviour is easily found to be,\footnote{ The only non--zero, relevant heat--kernel coefficient is the Weyl volume one, $C_0=2L$. The other, relevant coefficient, $C_{1/2}$, vanishes.} at fixed $\la$,
  $$\eqalign{
  \log F(m^2)&\sim 2mL+\log\bigg({\sinh^2a+\cosh^2b)\over \sinh^2a}\bigg)-2\log2+O(1/m)\cr
  \ze'(0,m^2)&\sim-2mL+O(1/m)\,.
  }
  $$
The dependence on the boundary eigenvalues, $\la$, is contained in the discarded $O(1/m)$ terms, which is the simplifying circumstance of this approach.

Applying (\peq{det}),
$$\eqalign{
    -\ze'(0,\la^2)&=\log F(\la^2)-\log\bigg({\sinh^2a+\cosh^2b)\over \sinh^2a}\bigg)+2\log2\cr
    &=-2b-\log\sinh^2a-\log\bigg({\sinh^2a+\cosh^2b)\over \sinh^2a}\bigg)+2\log2\cosh(\la L-b)\cr
    &=-2b-\log(\cosh a_0\cosh a_1)+2\la L +2\log\big(1+e^{2b-2\la L}\big)\,.
    }
    \eql{det3}
$$

It is necessary to take into account the contribution of the positive boundary modes, $\chi^{(+)}$, which, this time, is not equal to the $\chi^{(-)}$ one just calculated. It is obtained most easily by making the formal replacement $u\to-u^*$ in the modes, \cf\ (\peq{modes}), \ie $\la\to-\la$ and $k\to k$. This results in the replacement $b\to-b$ in (\peq{det3}) and adding the two gives,
$$\eqalign{
  -\ze_{tot}'(0,\la)  &=-2\log(\cosh a_1\cosh a_0)+4\la L \cr
  &\hspace{************}+2\log\big((1+e^{2b-2\la L})(1+e^{-2b-2\la L})\big)\,.
    }
    \eql{det4}
$$
Therefore, summing over $\la$, the total Dirac $\log\det$ on the finite cylinder is,
  $$\eqalign{
  \log\det D^2=&-2\log(\cosh a_0\cosh a_1)\,\ze_\man(0)+4 L\,\ze_\man(-1/2) \cr &  \hspace{**********}+2\sum_{\la>0}\log\big((1+e^{-2b-2\la L})(1+e^{2b-2\la L})\big)\,,
  }
  \eql{logdet3}
  $$
which, as a check, reduces to (twice) (\peq{logdet}) when $b=0$ (\ie $a_0=a_1, =a$) and is the final result of this section.

If, as here for spinors, the boundary $\ze_\man(0)$ vanishes, the only effect of the boundary condition is the introduction of a `chemical potential', $b/L$, \ie a constant shift in the boundary eigenvalues, positive and negative equally. This is not too surprising since chemical potentials are conventionally introduced through the quasi--periodicity of thermal Green functions, say.

\section{\bf 13. Higher dimensional spheres}
The preceding methods are not the only ones leading to the formula for the logdet, which is essentially a generalised first Kronecker limit formula. The usual way of deriving this is to apply Poisson summation \footnote{ Possibly through one of its many disguises such as a $\th$--function transformation.}  to an explicit eigenvalue form of the total \zf\ in the cylinder or, classically, on a torus, \eg [\pref{Weber2}] \S 115.

In the following, the discussion is confined to the half--integer (NS) case  as this has a conventional thermal aspect. (The integer case could be regarded as a kinematic spinor thermalised as a boson.) The other boundary conditions, are left for another time.

I outline the computation, [\pref{Apps,ApandD2}], of the functional determinant on $\gC$ which follows the same path as that laid down by Kronecker, Weber and  Epstein for $d=1$ leading to the Kronecker limit formula. The extension to higher dimensions is relatively straightforward so long as all the spectral data for S$^d$ are available. The best way of expressing these is to give the spinor \zf\ as a Barnes \zf,
  $$
  \ze_{S^d}^F(s)=\caS_d\,\ze_d(2s,d/2\mid{\bf1})\,,
  \eql{fzeta}
  $$
 $\caS_d$  is, more or less, the spinor dimension.

This effectively encodes the degeneracies and eigenvalues in a generating function and corresponds to a character representation in another language.

Applying simple (NS-type) mixed spinor boundary conditions on the cylinder, algebra then produces the intermediate formula,
    $$
    \ze_{I\times S^d}^F(s)={1\over2}r^{2s}\big[Z(s,2L)-Z(s,L)]
    $$
with the derivative,
   $$
       Z'(0,L)=-{2L\over r}\ze^F_{S^d}(-1/2)+2^{-(d-5)/2}\sum_{m=1}^\infty{1\over m}\cosech^d mL/r
   $$
so that the cylinder (one loop) effective action, $W^F$,   is, for odd $d$,
   $$
       W^F_{I\times S^d}\equiv{1\over2}{\ze^F_{S^1}}'(0)=-{L\over r}\ze^F_{S^d}(-1/2)+2^{-(d-1)/2}\sum_{m=1}^\infty{(-1)^m\over m}\cosech^d mL/r
       \eql{effact}
   $$

 The first term on the right--hand--side is proportional to the usual vacuum energy on the $d$--sphere,  obtained here from (\peq{fzeta}) as a generalised Bernoulli number, and the whole expression has the form of a thermodynamic free energy (multiplied by an inverse temperature).

 The even dimensional result is similar, except that there is no vacuum energy and the spin dimension is different. It is,
    $$
     W^F_{I\times S^d}=2^{1-d/2}\sum_{m=1}^\infty{(-1)^m\over m}\cosech^d mL/r\,,
       \eql{effact}
    $$
where the fact that the conformal anomaly is also zero for even dimensions has been used.\footnote{ This is a particular circumstance of this system.}

It is always helpful to have particular examples visible, [\pref{Apps}],
  $$\eqalign{
      W^F_{I\times S^1}&=\sum_{m=1}^\infty{(-1)^m\over m}\cosech mL/r-{L\over12r}\cr
       W^F_{I\times S^2}&=\sum_{m=1}^\infty{(-1)^m\over m}\cosech^2 mL/r\cr
        W^F_{I\times S^3}&={1\over2}\sum_{m=1}^\infty{(-1)^m\over m}\cosech^3 mL/r+{17L\over480r}\cr
         W^F_{I\times S^4}&={1\over2}\sum_{m=1}^\infty{(-1)^m\over m}\cosech^4 mL/r\cr
          W^F_{I\times S^5}&={1\over4}\sum_{m=1}^\infty{(-1)^m\over m}\cosech^5 mL/r-{367L\over24192r}\,.\cr
          }
       \eql{examples}
  $$

  The simple cylinder $W^F_{I\times S^1}$ has a maximum, as a function of $\mu=L/r$, at $\mu=\pi$ seen either as a consequence of its invariance under the inversion, $\mu\to \mu'=\pi^2/\mu$ or analytically by differentiating with respect to $\mu$ and using the summation \footnote{ This follows as a limiting case of the invariance formula so it's the same thing really. See Nanjundiah [\pref{Nand}].}
     $$
     \sum_{m=1}^\infty (-1)^m{\cosh m\pi\over\sinh^2m\pi}=-{1\over12}\,.
     $$

 An inspection of the signs shows that a maximum can be expected as $L/r$ varies only for the dimensions $d=1,5c,9,13,....$ . The sums converge rapidly and so these formulae are numerically adequate.

For completeness  the ellipticised simple cylinder case is, [\pref{dowdomains}],
   $$\eqalign{
        W^F_{I\times S^1}&=-2\log Q_2(q)-{L\over12}\cr
        &=-2\log \gf(q)
        }
   $$
where
  $$
        Q_2(q)=\prod_{n=1}^\infty (1+q^{2n-1})
  $$
and $\gf(q)$ is  a Weber class function.
\section{\bf 14. Conclusion}

After expressing various effective actions on the finite $2$-cylinder in terms of classic elliptic quantities, the mode structure of spinor theory, in $d+1$-dimensions, was analysed and the existence of real exponential modes along the cylinder for certain ranges of the parameters fixing these conditions was pointed out. The functional determinant for the most general bag boundary conditions was obtained.

Further computations might address the spectral asymmetry and also the nature of the charge conjugate to the chemical potential.
\section{\bf Acknowledgment}
I thank Klaus Kirsten for drawing my attention to the very useful reference [{\pref{KLP}] and for comments.
%\newpage

% ************************** REFERENCES ************************
\vskip 10truept
\noin{\bf{References}}
\vskip 5truept
\begin{putreferences}
   \ref{dowrobin}{J.S.Dowker, {\it Robin conditions on the Euclidean ball}, arXiv:hep-th/9506042,\break \cqg{13}{96}{585}.}
   \ref{Voros}{A.Voros, \cmp{110}{1987}{439}.}
     \ref{dowhybrid}{J.S.Dowker {\it The hybrid spectral problem and Robin boundary conditions}, arXiv: math/0409442, \jpa{38}{2005}{4735}.}
     \ref{EandK}{G.Esposito and K.Kirsten, {\it Chiral Bag Boundary Conditions on the Ball},
      arXiv: 0207109, \prD{66} {2002} {085014}.}
      \ref{FGKS}{G.Fucci, F.Gesztesy, K.Kirsten and J.Stanfill, {\it Spectral \zfs\ and $\ze$-- regularized functional determinants for regular Sturm-Liouville operators}, \break arXiv:2101.12295.}
     \ref{EGK}{G.Esposito, P.B.Gilkey and K.Kirsten {\it Heat Kernel Coefficients for Chiral Bag Conditions}, arXiv: math/0510156,  \jpa {38}{2005} {2259}.}
     \ref{BGKS}{C.G.Beneventano, P.B.Gilkey, K.Kirsten and E.M.Santangelo {\it Strong ellipticity and spectral properties of chiral bag boundary conditions}, arXiv: hep-th/0306156, \jpa{36}{2003}{11533}.}
    \ref{Strauss}{W.A.Strauss {\it Partial Differential Equations}, Wiley (New York) (2008).}
     \ref{BSW}{C.G.Beneventano, E.M.Santangelo and A.Wipf., {\it Spectral asymmetry for bag boundary conditions}, arXiv:hep-th/0205199, \jpa {35} {2002} 9343}
     \ref{Rayleigh}{Lord Rayleigh, {\it Note on the numerical calculation of the roots of fluctuating functions}, \plms{5}{1873}{119}.}
    \ref{Nand}{T.S.Nanjundia {\it Certain summations due to Ramanujan and their generalizations}, {\it Proc.Ind.Acad.Sci. Sect.A} {\bf 34} (1951) 215.}
    \ref{BandS}{C.G. Beneventano and E.M.Santangelo, {\it Finite--temperature properties of the Dirac operator under local boundary conditions}, \jpa {37}{2004}{9261}.}
   \ref{KLP}{K.Kirsten, P.Loya and J.Park, {\it Zeta functions of Dirac and Laplace--type operators over finite cylinders}, \aop{321}{2006}{1814}.}
   \ref{Cardy}{J.L.Cardy, \np{366}{91}{403}.}
   \ref{Luckock}{H.Luckock, {\it Mixed boundary conditions in quantum field theory}, \jmp{35}{91}{1755}.}
   \ref{dowdomains}{J.S.Dowker, {\it Casimir domains on a sphere}, arXiv:2105.11374 }
   \ref{Weber2}{H.Weber {\it Elliptische Functionen und Algebraische Zahlen}, Vieweg, (Braun-\break schweig) (1891).}
    \ref{BKM}{D.Burghelea, T.Kappeler and P.McDonald, {\it On the Functional logdet and Related Flows on the Space of Closed Embedded Curves on {\rm S}$^2$}, {\it J.Func.Anal} (1994) 440.}
     \ref{PandW}{E.Peltiola and Y.Wang {\it Large deviations of multichordal SLE0+, real rational functions and zeta-regularized determinants of Laplacians}, arXiv:2006.08574. }
     \ref{YiZ}{N.Yui and D.Zagier {\it Math. Comp} {\bf 66} (1997) 1645.}
     \ref{KSX}{H.Kleinert, E.Strobel and S-S.Xue {\it Phys. Rev.} D {\bf 88} (2013) 025049     .}
        \ref{DMS}{P.Di Francesco, P.Mathieu and D.S\'en\'echal {\it Conformal Field Theory}. Springer (New York) 1997.}
     \ref{Berndt2}{B.C.Berndt \jram{303/304}{1978}{332}.}
   \ref{SandWi}{N.Seiberg and E.Witten \np{276}{86}{272}.}
  \ref{dowtetra}{J.S.Dowker {\it Vacuum and magnetic field effects on a tetrahedron and other flat Riemann surfaces}, {\it Phys.Rev.} D{\bf{40}} (1989) {1938}.}
\ref{IZ}{C. Itzykson and J.B. Zuber,  \np{275}{86}{580}.}
  \ref{FKM}{G. Fucci, K. Kirsten and J.M. Mu\~noz-Caste\~neda {\it Casimir pistons with generalized boundary conditions: a step forward},  arXiv:1906.08486. }
\ref{DandK1}{J.S.Dowker,J.S. and K.Kirsten {\it Elliptic functions and
  temperature inversion symmetry on spheres}, \np {638} {2002} 405, arXiv:hep-th/0205029.}
  \ref{ZandR}{Zucker,I.J. and Robertson,M.M.
  {\it Math.Proc.Camb.Phil.Soc} {\bf 95 }(1984) 5.}
  \ref{JandZ1}{Joyce,G.S. and Zucker,I.J.
  {\it Math.Proc.Camb.Phil.Soc} {\bf 109 }(1991) 257.}
  \ref{JandZ2}{Zucker,I.J. and Joyce.G.S.
  {\it Math.Proc.Camb.Phil.Soc} {\bf 131 }(2001) 309.}
  \ref{zucker2}{I.J.Zucker {\it SIAM J.Math.Anal.} {\bf 10} (1979) 192.}
  \ref{BandZ}{Borwein,J.M. and Zucker,I.J. {\it IMA J.Math.Anal.} {\bf 12}
  (1992) 519.}
\ref{BandB}{J.M.Borwein  and P.B.Borwein, {\it Pi and the AGM}, Wiley,
  (New York) (1998).}
\ref{DandA}{Dowker,J.S. and Apps, J.S. \cqg{15}{1998}{1121}.}
 \ref{ApandD}{Apps,J.S. and Dowker,J.S. \cqg{15}{1998}{1121}.}
 \ref{ApandD2}{J.S.Dowker and J.S.Apps, {\it Further functional determinants},
 arXiv:hep-th \break /9502015, \cqg{12}{95}{1363}.}
\ref{DandA2}{J.S.Dowker and J.S.Apps, {\it Functional determinants on
certain domains}. {\it Int. J. Mod. Phys. D} {\bf 5} (1996) 799; hep-th/9506204.}
\ref{Apps}{J.S.Apps, PhD thesis (University of Manchester, 1996).}
\ref{Weber}{H.Weber, {\it Lehrbuch der Algebra, Bd.III}, Vieweg (Braunschweig) (1908).}
\ref{KCD}{G.Kennedy, R.Critchley and J.S.Dowker \aop{125}{80}{346}.}
\ref{Donnelly}{H.Donnelly \ma{224}{1976}161.}
\ref{Fur2}{D.V.Fursaev {\sl Spectral geometry and one-loop divergences on
manifolds with conical singularities}, JINR preprint DSF-13/94,
hep-th/9405143.}
\ref{HandE}{S.W.Hawking and G.F.R.Ellis {\sl The large scale structure of
space-time} Cambridge University Press, 1973.}
\ref{DandK}{J.S.Dowker and G.Kennedy \jpa{11}{78}{895}.}
\ref{ChandD}{Peter Chang and J.S.Dowker \np{395}{93}{407}.}
\ref{FandM}{D.V.Fursaev and G.Miele \pr{D49}{94}{987}.}
\ref{Dowkerccs}{J.S.Dowker \cqg{4}{87}{L157}.}
\ref{BandH}{J.Br\"uning and E.Heintze \dmj{51}{84}{959}.}
\ref{Cheeger}{J.Cheeger \jdg{18}{83}{575}.}
\ref{SandW}{K.Stewartson and R.T.Waechter \pcps{69}{71}{353}.}
\ref{CandW}{H.S.Carslaw and J.C.Jaeger {\it The conduction of heat
in solids}
Oxford, The Clarendon Press 1959.}
\ref{BandH}{H.P.Baltes and E.M.Hilf {\it Spectra of finite systems}.}
\ref{Epstein}{P.Epstein \ma{56}{1903}{615}.}
\ref{Kennedy1}{G.Kennedy \pr{D23}{81}{2884}.}
\ref{Kennedy2}{G.Kennedy PhD thesis, Manchester (1978).}
\ref{Kennedy3}{G.Kennedy \jpa{11}{78}{L173}.}
\ref{Luscher}{M.L\"uscher, K.Symanzik and P.Weiss \np {173}{80}{365}.}
\ref{Polyakov}{A.M.Polyakov \pl {103}{81}{207}.}
\ref{Bukhb}{L.Bukhbinder, V.P.Gusynin and P.I.Fomin {\it Sov. J. Nucl.
 Phys.} {\bf 44} (1986) 534.}
\ref{Alvarez}{O.Alvarez \np {216}{83}{125}.}
\ref{DandS}{J.S.Dowker and J.P.Schofield \jmp{31}{90}{808}.}
\ref{Dow1}{J.S.Dowker \cmp{162}{94}{633}.}
\ref{Dow2}{J.S.Dowker \cqg{11}{94}{557}.}
\ref{Dow3}{J.S.Dowker \jmp{35}{94}{4989}; erratum {\it ibid}, Feb.1995.}
\ref{Dow5}{J.S.Dowker {\it Heat-kernels and polytopes} To be published}
\ref{Dow6}{J.S.Dowker \pr{D50}{94}{6369}.}
\ref{Dow7}{J.S.Dowker \pr{D39}{89}{1235}.}
\ref{BandG}{T.P.Branson and P.B.Gilkey\tams{344}{94}{479}.}
\ref{Schofield}{J.P.Schofield Ph.D.thesis, University of Manchester, (1991).}
\ref{Barnesa}{E.W.Barnes {\it Trans. Camb. Phil. Soc.} {\bf 19} (1903)
374.}
\ref{BandG2}{T.P.Branson and P.B.Gilkey {\it Comm. Partial Diff. Equations}
{\bf 15} (1990) 245.}
\ref{Pathria}{R.K.Pathria {\it Suppl.Nuovo Cim.} {\bf 4} (1966) 276.}
\ref{Baltes}{H.P.Baltes \prA{6}{72}{2252}.}
\ref{Spivak}{M.Spivak {\it Differential Geometry} vols III, IV, Publish
or Perish, Boston, 1975.}
\ref{Eisenhart}{L.P.Eisenhart {\it Differential Geometry}, Princeton
University Press, Princeton, 1926.}
\ref{Moss}{I.Moss \cqg{6}{89}{659}.}
\ref{Barv}{A.O.Barvinsky, Yu.A.Kamenshchik and I.P.Karmazin \aop {219}
{92}{201}.}
\ref{Kam}{Yu.A.Kamenshchik and I.V.Mishakov {\it Int. J. Mod. Phys.}
{\bf A7} (1992) 3265.}
\ref{Death}{P.D.D'eath and G.V.M.Esposito, {\it Local boundary conditions for the Dirac operator and one--loop quantum cosmology}, \prD{43}{1991}{3234}.}
\ref{Rich}{K.Richardson \jfa{122}{94}{52}.}
\ref{Osgood}{B.Osgood, R.Phillips and P.Sarnak \jfa{80}{88}{148}.}
\ref{BCY}{T.P.Branson, S.-Y. A.Chang and P.C.Yang \cmp{149}{92}{241}.}
\ref{Vass}{D.V.Vassilevich.{\it Vector fields on a disk with mixed
boundary conditions} gr-qc /9404052.}
\ref{MandP}{I.Moss and S.Poletti \pl{B333}{94}{326}.}
\ref{Kam2}{G.Esposito, A.Y.Kamenshchik, I.V.Mishakov and G.Pollifrone
\prD{50}{94}{6329}.}
\ref{Aurell1}{E.Aurell and P.Salomonson \cmp{165}{94}{233}.}
\ref{Aurell2}{E.Aurell and P.Salomonson {\it Further results on functional
determinants of laplacians on simplicial complexes} hep-th/9405140.}
\ref{BandO}{T.P.Branson and B.\O rsted \pams{113}{91}{669}.}
\ref{Elizalde1}{E.Elizalde, \jmp{35}{94}{3308}.}
\ref{BandK}{M.Bordag and K.Kirsten {\it Heat-kernel coefficients of
the Laplace operator on the 3-dimensional ball} hep-th/9501064.}
\ref{Waechter}{R.T.Waechter \pcps{72}{72}{439}.}
\ref{GRV}{S.Guraswamy, S.G.Rajeev and P.Vitale {\it O(N) sigma-model as
a three dimensional conformal field theory}, Rochester preprint UR-1357.}
\ref{CandC}{A.Capelli and A.Costa \np {314}{89}{707}.}
\ref{IandZ}{C.Itzykson and J.-B.Zuber \np{275}{86}{580}.}
\ref{BandH}{M.V.Berry and C.J.Howls \prs {447}{94}{527}.}
\ref{DandW}{A.Dettki and A.Wipf \np{377}{92}{252}.}
\ref{Weisbergerb} {W.I.Weisberger \cmp{112}{87}{633}.}
%Reffs
\end{putreferences}
\newpage
\bye